\documentclass[prd,nofootinbib,showpacs,superscriptaddress,preprint]{revtex4}
\usepackage[T1]{fontenc}
\usepackage{amsmath,amssymb}
\usepackage{epsfig}
\usepackage{dcolumn}
\usepackage{graphicx}
\usepackage[usenames,dvipsnames]{color}
\usepackage{slashed}
\usepackage[colorlinks,citecolor=blue]{hyperref}
\usepackage{pdfpages}
\usepackage{float}
\usepackage{soul}

\begin{document}
	\title{ Neutrino mass, leptogenesis and sterile neutrino dark matter in inverse seesaw framework}
	
	\author{Nayana Gautam}
	\email{nayana@tezu.ernet.in}
	\affiliation{Department of Physics, Tezpur University, Tezpur - 784028, India}
	
	\author{Mrinal Kumar Das}
	\email{mkdas@tezu.ernet.in}
	\affiliation{Department of Physics, Tezpur University, Tezpur - 784028, India}
\begin{abstract}
	
We study $S_{4}$ flavor symmetric inverse seesaw model which has the possibility of simultaneously addressing neutrino phenomenology, dark matter (DM) and baryon asymmetry of the universe (BAU) through leptogenesis. The model is the extension of the standard model by the addition of two right handed neutrinos and three sterile fermions leading to a keV scale sterile neutrino dark matter and two pairs of quasi-Dirac states. The CP violating decay of the lightest quasi- Dirac pair present in the model generates lepton asymmetry which then converts to baryon asymmetry of the universe. Thus this model can provide a simultaneous solution for non zero neutrino mass, dark matter content of the universes, and the observed baryon asymmetry. The $S_{4}$ flavor symmetry in this model is augmented by additional $Z_{4}\times Z_{3}$ symmetry to constrain the Yukawa Lagrangian. A detailed numerical analysis has been carried out to obtain dark matter mass, DM-active mixing as well as BAU both for normal hierarchy as well as inverted hierarchy. We try to correlate the two cosmological observables and found a common parameter space satisfying the DM phenomenology and BAU. The parameter space of the model is further constrained from the latest cosmological bounds on the observables. 
\end{abstract}
\pacs{12.60.-i,14.60.Pq,14.60.St}
\maketitle	
\section{\label{sec:level1}Introduction}	
The existence of Dark Matter (DM), the baryon asymmetry of the universe (BAU), and the massive nature of neutrinos are significant observational evidence for physics beyond the standard model (SM). The extension of the standard model of particle physics can shed light on deciphering the nature of dark matter which has been a great challenge to the physics community worldwide \cite{mass-mixingneutrino,king2014neutrino}. Dark matter is considered to be a crucial ingredient of producing the large inhomogeneities that form structures in the universe. Cosmological and astrophysical observations in different context ensure the presence of dark matter \cite{rubin1970rotation,clowe2006direct}. According to the Planck data, 26.8 \% of the energy density in the universe is composed of DM and the present dark matter abundance is reported as  \cite{ade2016ade} 
\begin{equation*}
\Omega_{DM}h^2 = 0.1187 \pm 0.0017 
\end{equation*}
The requirements for a particle to be a good dark matter candidate as mentioned in \cite{taoso2008dark} are not fulfilled by any one of the SM particles. This has motivated the particle physics community to study different possible BSM frameworks that can give rise to the correct DM phenomenology and can also be tested at several different experiments \cite{mukherjee2017common}. Among different dark matter candidates, sterile neutrino is the most popular candidate which can give rise to the correct DM phenomenology \cite{adhikari2017white,kusenko2009sterile}. It is a right handed neutrino,singlet under the SM gauge symmetry having tiny mixing with the SM neutrinos and a long lifetime \cite{eVsterile}. This neutral long-lived and massive particle, especially in the keV range has opened the window towards the search for the particle nature of the dark matter. The most natural way to produce sterile neutrinos is by their admixture with the active neutrinos \cite{admixturesterile,barry2011light}. There are different frameworks for incorporating sterile neutrinos beyond SM. Inverse seesaw is among one of the scenarios where SM is extended by the addition of sterile neutrinos which is explained in later sections.  

Apart from the dark matter, an appealing mystery in particle physics, as well as cosmology is the matter-antimatter asymmetry of the universe also termed as baryon asymmetry of the universe (BAU): an excess of baryons over anti baryons. The qualitative measurement of the asymmetry can be expressed as the baryons to photon ratio \cite{PhysRevD.98.030001}
\begin{equation}
\eta_{B}= \frac{n_{B}- n_{\bar{B}}}{n_{\gamma}} = (6.04\pm0.08)\times10^{-10}
\end{equation}
This matter-antimatter asymmetry of the Universe has to be explained in terms of
a dynamical generation mechanism called baryogenesis which needs three important conditions: baryon number (B) violation, C and CP violation, the departure from thermal equilibrium as pointed out by Andrej Sakharov \cite{Sakharov:1967dj}. It has been realized that a successful model of baryogenesis cannot be attained within the standard model (SM) which insists on the necessity of a model beyond the SM. In this context, leptogenesis has been a very popular and interesting mechanism to explain the baryon asymmetry of the universe. As mentioned in the original proposal by Fukugita and Yanagida thirty years back, this mechanism satisfies all the Sakharov’s conditions required to produce the estimated baryon asymmetry \cite{Fukugita:1986hr}. In this mechanism, a net generated leptonic asymmetry gets converted into baryon asymmetry through B+L violating electroweak sphaleron transitions \cite{PILAFTSIS_1999,kolb2018early}. The interesting feature of this scenario is that the out of equilibrium decay of the same heavy fields responsible for producing neutrino mass and mixing can generate the required lepton asymmetry \cite{mukherjee2017common}. As outlined earlier, the mystery of dark matter can also be explainable in the seesaw framework. In this way, one can say that leptogenesis realizes a highly nontrivial link among three completely independent observations: matter-antimatter asymmetry, neutrinos mass and mixing, and also the dark matter of the Universe.

With the motivation of explaining baryon asymmetry of the universes and the dark matter within one framework, we have studied a special type of inverse seesaw (ISS) known as ISS(2,3) in which the particle content of SM is extended by two right handed(RH) neutrinos and three extra sterile fermions. In this framework, two RH neutrinos combine with two sterile neutrinos to form two quasi-Dirac states while one particle remains sterile which can lead to a keV scale sterile neutrino dark matter. Again, in the same framework, the lepton asymmetry is generated through the decay of the lightest quasi-Dirac pair present in the model. The model is constructed using discrete flavor symmetry $S_{4}$ and $Z_{4}$, $Z_{3}$ groups are further implemented to avoid unnecessary interactions among the particles. The matrices involved in ISS are constructed in such a way that the $\mu-\tau$ symmetry is broken in the resulting light neutrino mass matrix to comply with the experimental evidence of non-zero reactor mixing angle \cite{an2012observation}. The model parameters have been evaluated and using these parameters, we have carried out a phenomenological study of dark matter considering the lightest sterile neutrino as a potential candidate. Moreover, we feed the model parameters to the calculation of the lepton asymmetry generated by the lightest quasi-Dirac pair and found that this led to baryon asymmetry in agreement with the observed baryon asymmetry of the Universe. Thus our model is consistent with the point of baryon asymmetry as well as the dark matter of the Universe.

The paper is organized as follows. In section \ref{sec:level2}, we present the $S_{4}$  flavor symmetric ISS(2,3) model and the construction of different mass matrices in the lepton sector. Section \ref{sec:level3} is the brief discussion of leptogenesis in ISS(2,3) framework. In section \ref{sec:level4}, we briefly outline the keV scale sterile neutrino dark matter production and different constraints on it. The detailed numerical analysis and the results are discussed in section \ref{sec:level6}. Finally, we conclude in section \ref{sec:level7}.

\section{\label{sec:level2} $S_{4}$ flavor symmetric ISS(2,3) model}

Ample of previous works have been focused on different types of seesaw mechanisms to explain the smallness of neutrino mass and other related issues beyond standard model \cite{arhrib2010collider,foot1989see,mohapatra1981neutrino}. It is the extension of the standard model with extra singlet fermion having an advantage of lowering the mass scale to TeV range, which may be probed at LHC and in future neutrino experiments.

In inverse seesaw, the standard model is extended by the addition of right-handed ($N_{R}$) neutrinos and gauge singlet sterile neutrinos $s_{i}$. The relevant Lagrangian for the ISS is given as \cite{Abada2014},

\begin{equation}
L = - \dfrac{1}{2}{n_{L}^{T}C m n_{L}} + h.c
\end{equation}
where $C \equiv i\gamma^{2}\gamma^{0}$  is the charge conjugation matrix and 
$n_{L}=(\nu_{L,\alpha},\nu_{R,i}^{c},s_{j})^{T}$. Here, $\nu_{L,\alpha}$ with $\alpha= e$,$\mu$,$\tau$ are the SM left handed neutrino states and $\nu_{R,i}^{c}(i=1,\nu_{R})$ are right handed (RH) neutrinos,while $s_{j}(j=1,s)$ are sterile fermions. The neutrino mass matrix arising from this type of Lagrangian is of the form :
\begin{equation}
m = \left(\begin{array}{ccc}
0 &{M_{d}}^{T}& 0 \\
M_{d} & 0 & M_{N} \\
0 & M_{N}^{T} & \mu \\
\end{array}\right) 
\end{equation}
where $M_{d}$ ,$M_{N}$ and $\mu$ are complex matrices. Standard model neutrinos at sub-eV scale can be obtained from $M_{d}$ at electroweak scale, $M_{N}$ at TeV scale and $\mu$ at keV scale as explained in many literatures. 
Block diagonalisation of the above mass matrix leads to the effective light neutrino mass matrix as:
\begin{equation}
m_{\nu}\approx M_{d}(M_{N}^{T})^{-1}\mu M_{N}^{-1}M_{d}^{T}
\end{equation}
Depending on the number of fields in the model, a generic ISS realization is characterized by the following mass spectrum

1. Three light active states with masses of the form 
\begin{equation}
m_{\nu} = \mathcal{O}(\mu)\frac{k^{2}}{1+k^{2}}  , k = \frac{\mathcal{O}(M_{d})}{\mathcal{O}(M_{N})}
\end{equation}

2. \#$\nu_{R}$ pairs of pseudo-Dirac heavy neutrinos with masses$\mathcal{O}(M_{N})$ + $\mathcal{O}(M_{d})$.

3. \#s--\#$\nu_{R}$ light sterile states (present only if \#s \textgreater\#$\nu_{R}$) with masses $\mathcal{O}(\mu)$.

In this work, our framework is ISS(2,3) which is the extension of the SM by the addition of two RH neutrinos and three additional sterile fermions as mentioned \cite{Abada2014}. The motivation for using this special type of inverse seesaw is that besides accounting for the low energy neutrino data, it can also provide a viable dark matter candidate and the decay of the RH neutrinos in the model can produce the observed baryon asymmetry of the universe (BAU) \cite{abada2017neutrino}. We have used $S_{4}$ discrete flavor symmetry as our previous work \cite{gautam2019phenomenology}, however, the charge assignments and flavons are different leading to a different neutrino mass matrix. This is done to get a special texture of $M_{N}$ so that one can obtain the total Majorana mass matrix in a block diagonal form as shown in\ref{eq:4a} for successful leptogenesis. The implementation of an additional discrete symmetry $S_{4}$ would further constrain the model, enhancing its predictability, especially in relation to its flavor structure and CP properties. The structures of the different mass matrices $ M_{d}$, $M_{N}$ and $\mu$ originate explicitly from the flavor symmetry only. In this model, the lepton doublet, charged lepton singlet of the SM  in inverse seesaw model transform as triplet $3_{1}$ of $S_{4}$ while the SM singlet neutrinos $N_{R}$, SM Higgs doublet and sterile fermion transform as $2$ and  $1_{1}$ of $S_{4}$ respectively. Further $Z_{4}\times Z_{3}$ symmetry is imposed to get the desired mass matrix and to constrain the non-desired interactions of the particles. The particle content and the charge assignments are detailed in table \ref{tab1} , where in addition to the lepton sector and to the SM Higgs, flavons $\phi$, $\phi^{\prime}$,$\phi_{s}$, $\chi$, $\phi_{l}$,$\chi^{\prime}$ have been introduced.
\begin{table}
	\centering
	\begin{tabular}{|c|c|c|c|c|c|c|c|c|c|c|c|c|c|c|c|}
		
		\hline 
	Field	& $L$ & $l_{R}$ & $\bar{N_{R}}$ & H & s & $\phi$ &  $\phi^{\prime}$ &$\phi_{s}$ & $\phi_{l}$  & $\chi$ & $\chi^{\prime}$ \\ 
	\hline 
	$S_{4}$ &$3_{1}$  & $3_{1}$& $2$ & $1_{1}$ & $1_{1}$& $3_{2}$ & $3_{1}$ &$1_{1}$ & $1_{1}$  &$3_{2}$  & $3_{1}$   \\
	\hline 
	$SU(2)_{L}$ & $2$ & $1$ &$1$& $2$ & $1$ & $1$ &  $1$ &$1$&$1$ & $1$ & $1$  \\
	\hline 
	$Z_{4}$& $1$ & $1$ & $i$ &$1$& $+i$ & $-i$ & $-i$ &  $-1$ &$-i$ & $-i $ & $-i$\\
	\hline 
	$Z_{3}$& $\omega^{2}$ & $1$ &$1$& $1$ & $\omega$ & $\omega$ &  $\omega$ &$\omega$ & $\omega$ & $\omega^{2}$ & $\omega^{2}$  \\
	\hline 
	\end{tabular} 
	\caption{Fields and their respective transformations under the symmetry group of the model.} \label{tab1}
\end{table}
The Yukawa Lagrangian for the charged leptons and also for the neutrinos can be expressed as:
\begin{equation}\label{eq:2}
-\mathcal{L}  = \mathcal{L}_{\mathcal{M_{L}}}+\mathcal{L}_{\mathcal{M_{D}}} + \mathcal{L}_{\mathcal{M}}+ \mathcal{L}_{\mathcal{M_{S}}}+ h.c
\end{equation}
where,
\begin{equation}\label{eq:3}
\mathcal{L}_{\mathcal{M_{D}}} =  \frac{y}{\Lambda}\bar{N_{R}}LH\phi +  \frac{y\prime}{\Lambda}\bar{N_{R}}LH\phi^{\prime},
\end{equation}
The cut-off scale $\Lambda$ is needed to lower the mass dimension to $4$. Here, we have also taken extra scalar $\phi$ and $\phi^{\prime}$ with SM Higgs because $S_{4}$ product between a doublet and triplet yields two triplets and H is a $S_{4}$ singlet, so we need another triplet scalar to keep the Lagrangian singlet under $S_{4}$.
\begin{equation}
\mathcal{L}_{\mathcal{M_{S}}} = y_{s}{\bar{{s_{1}}^{c}}}s_{1}\phi_{s}+ y_{s}{\bar{{s_{2}}^{c}}}s_{2}\phi_{s}+ y_{s}{\bar{{s_{3}}^{c}}}s_{3}\phi_{s}
\end{equation}
$ \mathcal{L}_{\mathcal{M_{L}}}$ is the Lagrangian for the charged leptons which can be written in terms of dimension five operators as
\begin{equation}
\mathcal{L}_{\mathcal{M_{L}}} = \frac{y_{l}}{\Lambda}\bar{L}l_{R}H\phi_{l} 
\end{equation}
The following flavon alignments allow us to have the desired mass matrix corresponding to the charged lepton sector 
$$\langle \phi_{l} \rangle = v_{l}.$$

The charged lepton mass matrix is then given by,
\begin{equation}\label{eq:g}
m_{l}= \frac{v_{h}}{\Lambda}\left(\begin{array}{ccc}
y_{l}v_{l} & 0& 
0 \\
0 & y_{l}v_{l} &
0\\ 
0& 0 & y_{l}v_{l}
\end{array}\right),
\end{equation}
The Lagrangian $\mathcal{L}_{\mathcal{M_{N}}}$ is given by,
\begin{equation}
\mathcal{L}_{\mathcal{M}} = \frac{y_{1r}}{\Lambda} {(\bar{N_{R}}\chi\chi^{\prime})}_{1}s_{1} + \frac{y_{2r}}{\Lambda} {(\bar{N_{R}}\chi^{\prime}\chi^{\prime})}_{1}s_{2} +\frac{y_{3r}}{\Lambda} {(\bar{N_{R}}\chi\chi)}_{1}s_{3}
\end{equation} 
The VEV alignments for the flavons which result in desired neutrino mass matrix and leptonic mixing matrix are followed as:
$$\langle \phi \rangle =v_{h1}(1,-1,1), \; \langle \phi^{\prime} \rangle = v_{h1^{\prime}}(-1,1,1), \;\langle \phi_{s} \rangle = v_{s}$$, 
$$\langle H \rangle =  v_{h},\; \langle\chi \rangle = v_{r}(1,0,0),\; \langle\chi^{\prime} \rangle = v_{r}^{\prime}(0,1,0) 
$$
These VEV alignments are allowed under $S_{4}$ flavor symmetry \cite{Krishnan:2012me}. With these flavon alignments different matrices involved in the model can be written as;
\begin{equation}
M_{d} = \frac{v_{h}}{\Lambda}\left(\begin{array}{ccc}

-y v_{h1} + y v_{h1^{\prime}} &  y v_{h1} - y v_{h1^{\prime}}& y v_{h1} - y v_{h1^{\prime}}  \\
y v_{h1} + y v_{h1^{\prime}} & -y v_{h1} - y v_{h1^{\prime}}& y v_{h1} + y v_{h1^{\prime}}  \\
\end{array}\right).
\end{equation}
\begin{equation}
M_{N} = \frac{1}{\Lambda}\left(\begin{array}{ccc}
y_{1r} v_{r}v_{r}^{\prime} & 0 & 0 \\
0 & y_{2r}v_{r}^{\prime}v_{r}^{\prime} & 0\\
\end{array}\right).
\end{equation}
\begin{equation}
\mu  = y_{s}\left(\begin{array}{ccc}
1 & 0 & 0 \\
0 & 1 & 0\\ 
0 & 0 & 1
\end{array}\right)v_{s}
\end{equation}
Now,we denote  $a =  \frac{v_{h}}{\Lambda}(-y v_{h1} + y v_{h1^{\prime}})$,  $b = \frac{v_{h}}{\Lambda}(y v_{h1} - y v_{h1^{\prime}})$,  $c = \frac{v_{h}}{\Lambda}(y v_{h1} + y v_{h1^{\prime}})$,  
$e = \frac{v_{h}}{\Lambda}( -y v_{h1} - y v_{h1^{\prime}})$, $f =\frac{1}{\Lambda}y_{1r} v_{r}v_{r}^{\prime}$, $g =\frac{1}{\Lambda} y_{2r} v_{r}^{\prime}v_{r}^{\prime}$, $p =y_{s} v_{s}$. With these notations the mass matrices involved in ISS can be written as,
\begin{equation} \label{eq:u} 
M_{d}= \left(\begin{array}{ccc}
a & b & b \\
c & e & c \\ 
\end{array}\right),\;  \mu  = \left(\begin{array}{ccc}
p & 0 & 0 \\
0 & p & 0\\ 
0 & 0 & p
\end{array}\right), \;  M_{N} =\left(\begin{array}{ccc}
f & 0 & 0 \\
0 & g & 0 
\end{array}\right).
\end{equation}

Thus the model with $S_{4}$ flavor symmetry leads to mass matrices different from our previous work \cite{gautam2019phenomenology} resulting in different dark matter phenomenology.

The mass matrix in ISS framework with $\|\mu\|\leqslant\|M_{d}\|,\|M_{N}\|$ can be block diagonalised into light and heavy sectors as \cite{awasthi2013neutrinoless}
\begin{equation}\label{eq:a}
m_{\nu}\approx M_{d}(M_{N}^{T})^{-1}\mu M_{N}^{-1}M_{d}^{T} 
\end{equation}
\begin{equation}
M_{H}= \left(\begin{array}{cc}
0 & M_{N} \\
{M_{N}}^{T} & \mu \\
\end{array}\right).
\end{equation}
where $m_{\nu}$ is the ISS formula and $M_{H}$ is the mass matrix for the heavy quasi-Dirac pairs and the extra state. $m_{\nu}$ is diagonalised by PMNS $U_{\nu}$ to get the light active neutrinos. While the diagonalisation of $M_{H}$ will give the mass of the other five heavy particles.\\ The active-sterile mixing in the framework of ISS as is given as \cite{lindner2014neutrino},
\begin{equation}\label{eq:e}
\epsilon\approx \dfrac{1}{2} M_{d}^{\dagger}({{M_{N}}^{-1}})^{\ast}({M_{N}}^{T})^{-1} M_{d}
\end{equation}
As mentioned in \cite{lindner2014neutrino} from equation \eqref{eq:e} and \eqref{eq:a} one may write, 
\begin{equation}\label{eq:f}
m_{\nu}\approx M_{d}(M_{N}^{T})^{-1}\mu M_{N}^{-1}M_{d}^{T}\approx \mu\epsilon  
\end{equation}
or 
\begin{equation}\label{eq:h}
\epsilon \approx m_{\nu} \mu^{-1}  
\end{equation}
In the framework of ISS(2,3), $M_{N}$ is not a squared matrix rather it is a $2\times3$ matrix. So, $M_{N}^{-1}$ is not well defined. We followed the general version of \eqref{eq:a} . It follows as,
\begin{equation}\label{eq:20}
m_{\nu}\approx M_{d}^{T}dM_{d} 
\end{equation} 
where d is $2\times2$ dimensional sub matrix defined as
\begin{equation}
{M_{H}}^{-1} = \left(\begin{array}{cc}
d_{2\times2} & .... \\
..... &.... \\
\end{array}\right).
\end{equation}
with 
\begin{equation}
M_{H}= \left(\begin{array}{cc}
0 & M_{N} \\
{M_{N}}^{T} & \mu \\
\end{array}\right).
\end{equation}
Similarly, the active-sterile mixing can be obtained as,
\begin{equation}\label{eq:b}
\epsilon\approx \dfrac{1}{2}M_{d}^{T}dM_{d}\mu^{-1}.
\end{equation}
The elements of the light neutrino mass matrix in the framework of ISS(2,3) arising from the above mentioned mass matrices are given below:
\begin{equation}
(-m_{\nu})_{11} = \frac{a^{2}p}{f^{2}}+\frac{c^{2}p}{g^{2}}
\end{equation}
\begin{equation}
(-m_{\nu})_{12} =\frac{abp}{f^{2}}+\frac{cep}{g^{2}}
\end{equation}
\begin{equation}
(-m_{\nu})_{13} = \frac{abp}{f^{2}} + \frac{c^{2}p}{g^{2}} 
\end{equation}
\begin{equation}
(-m_{\nu})_{22} =\frac{b^{2}p}{f^{2}}+\frac{e^{2}p}{g^{2}}
\end{equation}
\begin{equation}
(-m_{\nu})_{23} = \frac{b^{2}p}{f^{2}}+\frac{cep}{g^{2}}
\end{equation}
\begin{equation}
(-m_{\nu})_{33} = \frac{b^{2}p}{f^{2}}+\frac{c^{2}p}{g^{2}}
\end{equation}

The light neutrino mass matrix obtained here in both the models can give rise to the correct mass squared difference and non-zero $\theta_{13}$. Thus the desired structures of the mass matrices have been made possible by the combination of flavor symmetry $S_{4}$ as well as $Z_{4}\times Z_{3}$ symmetry.

\section{\label{sec:level3}Leptogenesis in inverse seesaw framework}
As mentioned above, the ISS(2,3) model contains three active neutrinos, two right handed neutrinos, and three extra right handed sterile fermions. In this framework, the four right handed neutrinos will combine to form two quasi-Dirac pairs \cite{awasthi2013neutrinoless} while one remains as sterile \cite{abada2014dark}. This framework has the advantage that the RH neutrinos present in the model can be responsible for the generation of the observed baryon asymmetry. The decay of the heavy Majorana particles will create lepton asymmetry which in turn can be converted to baryon asymmetry by the sphaleron process \cite{Lucente:2018uaj,Hambye_2012,Choubey_2010,Agashe:2018cuf}. In our model, the decay of the lightest quasi- Dirac pair will potentially contribute to the observed baryon asymmetry of the universe (BAU). The lepton asymmetries produced in the decay of heavier pair will be washed out by the lepton number violating scatterings of the lightest pair \cite{dev2010tev}. Thus the inverse seesaw has the advantage that it can explain the small mass of the active neutrinos with low scale RH neutrinos as well as the decay of the same RH neutrinos will produce the observed baryon asymmetry. It is to note that all the Dirac Yukawa couplings coming from the inverse seesaw are complex and hence can act as a source of CP-violation, as there is no CP violating phase associated with RH neutrinos \cite{borah2018common}. 

In the model, there are two quasi-Dirac RH neutrino pairs $N_{i,j}$ with masses $M_{i,j}$ where $M_{i}(i=1,2...4)$ denotes the four heavy neutrino mass eigenvalues while the fifth one corresponds to the lightest sterile neutrino \cite{gautam2019phenomenology}. The CP asymmetry generated by the decay of $N_{i}$ into any lepton flavor can be obtained as \cite{Covi_1996},
\begin{equation}\label{50}
\epsilon_{i} = \frac{1}{8\pi}\sum_{j\neq i}\frac{Im[(hh^{\dagger})_{ij}^{2}]}{\sum_{\beta}|h_{i\beta}|^{2}}f_{ij}^{\nu}
\end{equation}
where, $h_{i\alpha}$ represents the effective Yukawa coupling in the diagonal mass basis. $f_{ij}^{\nu}$ is given as \cite{Dev_2015},
\begin{equation}\label{51}
f_{ij}^{\nu} \simeq \frac{(M_{j}^{2}-M_{i}^{2})}{(M_{j}^{2}-M_{i}^{2})^{2}+(M_{j}\Gamma_{j}-M_{i}\Gamma_{i})^{2}}
\end{equation}
Here, $\Gamma_{i}$ is the decay width of the heavy-neutrino $N_{i}$.
$M_{i}$ are the real and positive eigenvalues of the heavy neutrino mass matrix which are grouped into two quasi-Dirac pairs with the mass splitting of order $\mu_{kk}$ $(k=1,2)$ while the remaining one will be the lightest sterile fermion. 
The calculation of CP asymmetry in terms of the Yukawa couplings in the flavor basis can be obtained using the relations among the Yukawa couplings in the flavor basis $(y_{i\alpha})$ and that in the mass basis $(h_{i\alpha})$ given in appendix \ref{appen1}. Using these relations, the expressions for CP asymmetry for the decay of one of the quasi-Dirac particles (say i=1) can be written as \cite{Blanchet_2010},
\begin{equation}\label{55}
\epsilon_{1} = \frac{1}{8\pi}\sum_{j\neq 1}\frac{Im[(hh^{\dagger})_{1j}^{2}]}{\sum_{\beta}|h_{1\beta}|^{2}}f_{1j}^{\nu}
= \frac{\varepsilon_{2}}{{16 \pi \sum_{\beta}|y_{1\beta}|^{2}}}Im[e^{i(\theta_{1}-\theta_{2})}(\sum_{\alpha}y_{1\alpha}^{*}y_{2\alpha})^{2}]f_{13}^{\nu}
\end{equation}
Here, we have taken $f_{13} = f_{14}$ and j=2 term vanishes as there is no imaginary part.
The term $e^{i\theta_{i}}$ comes from unitary transformation of a new matrix $\tilde{M_{i}}$ obtained from $M_{H}$ as shown here. The formation of the new matrix can be found in appendix as mentioned in \cite{Blanchet_2010}.
In the $(N_{i}, s_{i})$ flavor basis, we have the $2\times2$ matrices as,
\begin{equation}
\tilde{M_{i}}= \left(\begin{array}{cc}
0 & M_{N_i} \\
M_{N_i}&\mu_{ii} \\
\end{array}\right) =
\left(\begin{array}{cc}
0 & M_{N_i} \\
M_{N_i}& \varepsilon_{i}M_{N_i}e^{i\theta_{i}} \\
\end{array}\right).
\end{equation}
Here, $\varepsilon_{i} = \frac{\mu_{ii}}{M_{Ni}}<< 1$.Now, the wash out parameter $K_{i}$ are defined as 
\begin{equation}\label{eq:1a}
K_{i} = \frac{\Gamma_{i}}{H(z=1)} = \frac{M_{i}}{8\pi}(hh^{\dagger})_{ii}\times\frac{M_{pl}}{1.66\sqrt{g*} M_{1}^{2}}
\end{equation}
where, $\Gamma_{i}$ is the decay width of the decaying right handed neutrino (RHN) $N_{i}$ and H is the Hubble rate of expansion at temperature $T = M_1$ is $H= 1.66 \sqrt{g_*}\frac{M_{i}^2}{M_{pl}}$ We denote the effective number of relativistic degrees of freedom as $g_*$ and is approximately 110. Again, the decay width of one of the quasi-Dirac pairs (say i=1) can be given as ,
\begin{equation}\label{64}
\Gamma_{i} = \frac{M_{i}}{8\pi}(hh^{\dagger})_{ii}
\end{equation} 
From the expression above for lepton asymmetry, one can write the final BAU as,
\begin{equation}\label{eq:bau}
Y_{B} = c \sum_{i}{\kappa}_{i} {\epsilon}_{i}
\end{equation}
where $c$ determines the fraction of lepton asymmetry being converted into baryon asymmetry, the value of $c$ is $10^{-2}$. $\kappa$ is the dilution factor responsible for the wash out processes which erase the generated asymmetry. ${\epsilon}_{i}$ represents the CP asymmetry generated by the decay of RH neutrinos $N_{i}$ into any lepton flavor.  

The expressions for $\kappa$ in Eq.\ref{eq:bau} depending on the scale of the wash out factor $K$ \cite{mukherjee2018normal}.
\begin{gather}\label{eq:washout}
-\kappa \approx \sqrt{0.1K} \text{exp}[\frac{-4}{(3(0.1K)^{0.25})}],\hspace{6mm} \text{for} \hspace{6mm}    K \geq 10^6\\
\approx \frac{0.3}{K(\text{ln} K)^{0.6}}, \hspace{6mm} \text{for}  \hspace{6mm}        10 \leq K \leq 10^6\\
\approx \frac{1}{2\sqrt{K^2+9}},\hspace{6mm} \text{for}  \hspace{6mm}    0\leq K \leq 10.
\end{gather}
In our model, CP asymmetry generated by the heavy pair is washed out. Therefore, the CP asymmetry generated by the light quasi-Dirac pair can only have significant contributions to baryon asymmetry. Moreover, flavor effects are also not important in our framework which is an advantage of inverse seesaw because of the large Yukawa couplings as mentioned \cite{Blanchet_2010}. The analytical expressions of all the terms involved in the calculation in terms of our model parameters are explained in appendix.
\subsection*{\textbf Effect of $\Delta L=2$ Scattering processes on washout parameter:}
As explained in literature \cite{Pilaftsis_2005,Pilaftsis_1997,Giudice_2004}, $\Delta L=2$ scatterings mediated by on shell heavy neutrinos $l_{\alpha}\phi^{\dagger}\longrightarrow \bar{l_{\beta}}\phi$ can be used to study the interaction properties of heavy RH neutrinos. As mentioned  by the author in  \cite{Blanchet:2009kk}, the $\Delta L=2$ scattering can be separated into on shell and off shell contributions $\gamma_{\Delta L =2,\alpha}^{tot}=\gamma_{\Delta L =2,\alpha}^{onshell}+ \gamma_{\Delta L =2,\alpha}^{offshell} $. The significance of the on-shell part of the  $\Delta L=2$ scattering is that it provides the inverse decays of leptons into heavy neutrinos and can be expressed as \cite{Blanchet:2009kk},
\begin{equation}
\gamma_{\Delta L =2,\alpha}^{onshell}= \frac{\gamma_{N_{i},\alpha}^{D}}{4}=\frac{n_{N_{i}}^{eq}}{4}\frac{\mathcal{K}_{1}}{\mathcal{K}_{2}}\Gamma_{N_{i},\alpha} =\frac{4}{z}\frac{M_{i}^{4}}{128 \pi^{3}}|h_{i\alpha}|^{2}\mathcal{K}_{1}(z M_{i}/M_{1})
\end{equation}	
where $\mathcal{K}_{1,2}(z)$ are the modified Bessel functions. Inverse decays may deplete the lepton asymmetry, parametrized by the washout parameter. From the above expression, it is clear that the on shell contribution does not vanish even at L conserving limits giving a non zero value of the$\Delta L=2$ scatterings which is contradictory to the fact that in absence of Lepton number violation $\Delta L=2$ scattering must vanish. As shown by the authors \cite{Blanchet:2009kk}, this problem can be solved by considering the interference terms along with the inverse decay terms in the on shell contribution. One can see \cite{Blanchet:2009kk} for the detailed calculation. Adding the interference term, the wash out parameter gets modified to,
\begin{equation}\label{eq:1b}
K_{\alpha}^{eff} \simeq K_{\alpha}.\delta^2
\end{equation}
Where $\delta = \frac{M_{2}-M_{1}}{\Gamma_{1}} = \frac{\mu_{11}}{\Gamma_{1}} $.
Thus the effective wash parameter is suppressed in presence of such a scattering process. Though K in Eq.\ref{eq:1a} is large in TeV Scale, yet it is suppressed by the factor $\delta$ ($\delta<<1$) as shown in Eq.\ref{eq:1b}. $\delta$ is very small ($\sim 10^{-5}-10^{-4}$) as it is proportional to $\mu_{11}$ which must be very small to get the scale of the light neutrinos. In our work, we use $K_{\alpha}^{eff}$ to calculate the baryon asymmetry.

\section{\label{sec:level4}Sterile neutrinos in keV scale}
Another important feature of our model is that it can naturally lead to a sterile state besides explaining the baryon asymmetry of the universe (BAU). This state can have any masses from eV to MeV depending on the model. Our study includes sterile neutrino specifically in the keV scale which has both phenomenological and cosmological importance. Sterile neutrinos in the keV range can be a viable candidate to account for the dark matter (DM) of the universe \cite{lucente2016implication}. 

Sterile neutrino being a neutral, massive and long lived particle can be a good dark matter candidate \cite{Merle2017a,abazajian2017sterile}. However, sterile neutrino must satisfy the bounds from astrophysics and cosmology on the mass and mixing to fulfill the criteria of being DM candidate. If sterile neutrino is a DM candidate, the relic abundance is proportional to the active-sterile mixing and the mass and can be expressed as \cite{abada2014dark,asaka2007lightest}:
\begin{equation}\label{eq:c}
\Omega_{DM}h^{2} = 1.1 \times 10^{7}\sum C_{\alpha}(m_{s})|\epsilon_{\alpha s}|^{2}{\left(\frac{m_{s}}{keV}\right)}^{2},  \alpha = e,\mu,\tau
\end{equation}
where $sin^{2}2\theta = 4 \sum|\epsilon_{\alpha s}|^{2}$ with $|\epsilon_{\alpha s}|$is the active-sterile leptonic mixing matrix element and $m_{s}$ represents the mass of the lightest sterile fermion.

One of the important criteria for a DM candidate is its stability on cosmological scale. The lightest sterile neutrino may decay into an active neutrino and a photon $\gamma$ via the process $N\longrightarrow\nu+\gamma$ that leads to a monochromatic X-ray line signal \cite{perez2017almost}. However, as discussed in many literature, the decay rate is negligible with respect to the cosmological scale because of the small mixing angle. The decay rate is given as \cite{ng2019new} :
\begin{equation}\label{eq:d}
\Gamma=1.38\times10^{-32}{\left(\frac{sin^{2}2\theta}{10^{-10}}\right)}{\left(\frac{m_{s}}{keV}\right)}^{5}s^{-1}.
\end{equation}
It is evident from the above expressions that the decay rate and as well as the relic abundance depend on mixing and mass of the DM candidate. 

Sterile neutrinos produced by the mixing with active neutrinos can have impacts on several phenomena like lepton number violating processes, deviation from unitarity, perturbativity of Yukawa couplings as mentioned in \cite{abada2016impact}. We study all the possible bounds on the sterile neutrino proposed in our model.

To ensure the compatibility with the latest neutrino oscillation parameters, we have evaluated the model parameters using the latest global fit neutrino oscillation data \cite{de2018status} and then use these numerically evaluated model parameters to study DM phenomenology. Again, in our analysis, Yukawa couplings lie in the range $(10^{-5}-10^{-2})$ which satisfy the perturbativity limits. Several cosmological observations also constrain the sum of the three neutrino masses for both normal and inverted hierarchy \cite{Aghanim:2018eyx} in presence of extra heavy sterile state. The proposed model leads to sum of the active neutrino masses within the cosmological range.

Addition of extra sterile states may violate the unitarity of the PMNS matrix. However, in our analysis, it did not affect the active neutrino mixing matrix (PMNS matrix). In our framework, the effective light neutrino mass matrix $m_{\nu}$ can be written as
\begin{equation}
m_{\nu} = U^{*}{m_{\nu}^{diag}}U^{\dagger}
\end{equation}
The unitary matrix U is related to the PMNS mixing matrix $U^{\nu}$ as follows \cite{abada2017neutrino}
\begin{equation}
U^{\nu} = (1-\frac{1}{2}\theta {\theta}^{\dagger})U + \mathcal{O}(\theta^{3})
\end{equation}
As mentioned by the authors in \cite{abada2017neutrino}, strong experimental constraints allow us to neglect $\theta$ which parametrises the deviation from unitarity of the PMNS matrix.

Direct detection and indirect detection of sterile neutrino provide strong bounds on the parameter space of mass and mixing with the active neutrinos. We have restricted our model with the bounds from direct detection experiments like XENON100, XENON1T \cite{campos2016testing}. Bounds from indirect detection come from satellite detectors like CHANDRA \cite{horiuchi2014sterile} and XMN \cite{boyarsky2008constraints}. We have incorporated the constraints on the mass-mixing parameter space as mentioned in \cite{abazajian2012light,perez2017almost,ng2019new}.

Sterile neutrinos produced via Dodelson-Widrow (DW) mechanism are warm DM, and constraints exist on their relic density from large scale structure (LSS) formation. In particular, the new limits on the nature of DM comes from Lyman-$\alpha$ forest data \cite{meiksin2009publisher} which constrain the mass-mixing parameter space. Constraints from LSS provide lower bounds on the DM mass \cite{boyarsky2009realistic}. In our model, sterile neutrinos are produced via non-resonant production (NRP) mechanism. Constraints from Lyman-$\alpha$ data on the mass of the sterile neutrino have been considered in our analysis. We have adopted the results in \cite{baur2017constraints,boyarsky2009lyman}. In this work, we have considered XQ-100 Ly-$\alpha$ data which is also compatible with SDSS-I + UVES data.\cite{Baur_2016,Y_che_2017}

\section{\label{sec:level6}Numerical Analaysis and Results}
The light neutrino mass matrix is diagonalised by a unitary PMNS matrix as,
\begin{equation}\label{eq:16}
m_{\nu} = U_{\text{PMNS}}m^{\text{diag}}_{\nu} U^T_{\text{PMNS}}
\end{equation}
where the Pontecorvo-Maki-Nakagawa-Sakata (PMNS) leptonic mixing matrix can be parametrized as \cite{giganti2017neutrino}
\begin{equation}
U_{\text{PMNS}}=\left(\begin{array}{ccc}
c_{12}c_{13}& s_{12}c_{13}& s_{13}e^{-i\delta}\\
-s_{12}c_{23}-c_{12}s_{23}s_{13}e^{i\delta}& c_{12}c_{23}-s_{12}s_{23}s_{13}e^{i\delta} & s_{23}c_{13} \\
s_{12}s_{23}-c_{12}c_{23}s_{13}e^{i\delta} & -c_{12}s_{23}-s_{12}c_{23}s_{13}e^{i\delta}& c_{23}c_{13}
\end{array}\right) U_{\text{Maj}}
\label{matrixPMNS}
\end{equation}
where $c_{ij} = \cos{\theta_{ij}}, \; s_{ij} = \sin{\theta_{ij}}$ and $\delta$ is the leptonic Dirac CP phase. The diagonal matrix $U_{\text{Maj}}=\text{diag}(1, e^{i\alpha}, e^{i(\beta+\delta)})$  contains the Majorana CP phases $\alpha, \beta$.
The diagonal mass matrix of the light neutrinos can be written  as, $m^{\text{diag}}_{\nu} 
= \text{diag}(0, \sqrt{m^2_1+\Delta m_{solar}^2}, \sqrt{m_1^2+\Delta m_{atm}^2})$ for normal hierarchy and  $m^{\text{diag}}_{\nu} = \text{diag}(\sqrt{m_3^2+\Delta m_{atm}^2}, 
\sqrt{\Delta m_{solar}^2+ \Delta m_{atm}^2},0)$ for inverted hierarchy \cite{Nath_2017}.
\begin{table}[H]
	\centering
	\begin{tabular}{|c|c|c|}
		
		\hline 
		Oscillation parameters	& 3$\sigma$(NO) & 3$\sigma$(IO) \\ 
		\hline 
		$\frac{\Delta m_{21}^{2}}{10^{-5}eV^{2}}$	& 7.05 - 8.14 &7.05 - 8.14  \\ 
		
		$\frac{\Delta m_{31}^{2}}{10^{-3}eV^{2}}$	& 2.41 - 2.60 &  2.31-2.51 \\ 
		
		$sin^{2}\theta_{12}$ &0.273 - 0.379  & 0.273 - 0.379 \\ 
		
		$sin^{2}\theta_{23}$ &  0.445 - 0.599  & 0.453 - 0.598 \\ 
		
		$sin^{2}\theta_{13}$ &  0.0196 - 0.0241 &  0.0199 - 0.0244 \\ 
		
		$\frac{\delta}{\pi}$ & 0.87 - 1.94 &  1.12- 1.94\\ 
		\hline 
	\end{tabular} 
	\caption{Latest Global fit neutrino oscillation Data \cite{de2018status}.}\label{tab3}
\end{table} 

For the numerical analysis, we have fixed the value of two model parameters f and g corresponding to the RH neutrino mass matrix, and we vary the model parameter p in a range so that it can lead to sterile neutrinos in the keV range as well as explain the baryon asymmetry of the Universe. The values of f and g are fixed at $9\times10^4$ GeV and $13.5\times10^4$ GeV respectively. The other four model parameters can be numerically evaluated by comparing the neutrino mass matrix arising from the model with the one which is parametrized by the available $3\sigma$ global fit data given in table \ref{tab3}. To ensure the equality of the other two elements, a tolerance of $10^{-4}$ is chosen. It means that we have taken those sets of model parameters for which the differences between the remaining two elements of the neutrino mass matrix arising from the model and the one which is parametrized by the available $3\sigma$ global fit data are less than $10^{-4}$. It has been observed that these two constraints tightly restrict the light neutrino parameters to a set of very specific values, resulting in a very predictive scenario.

With the same set of model parameters, numerically evaluated for the model, we have performed the calculations in lepton asymmetry using the equations mentioned above. We have taken the lightest sterile neutrino in the keV range to obtain dark matter phenomenology. We then calculate the baryon asymmetry for the light neutrino parameters that are consistent with neutrino data as well as the model restrictions discussed above. In the calculations, we have considered the decay of the lightest quasi-Dirac pair $(N_{1},s_{1})$ as the asymmetry generated by the heavy pair $(N_{2},s_{2})$ is washed out very rapidly.

\begin{figure}[H]
	\begin{center}
		\includegraphics[width=0.30\textwidth]{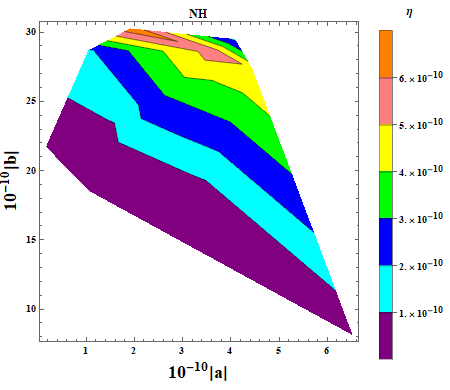}
		\includegraphics[width=0.30\textwidth]{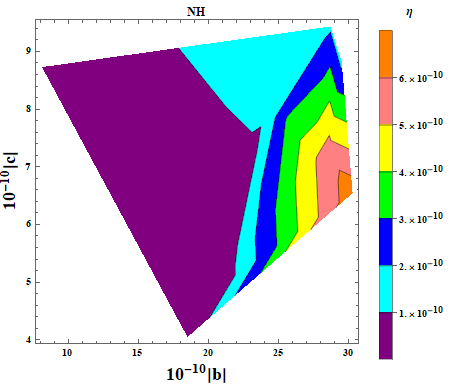}\\ 
		\includegraphics[width=0.30\textwidth]{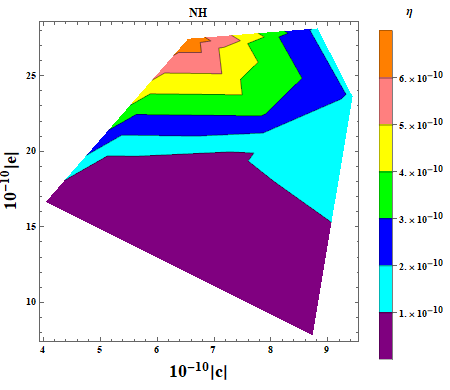}
		\includegraphics[width=0.30\textwidth]{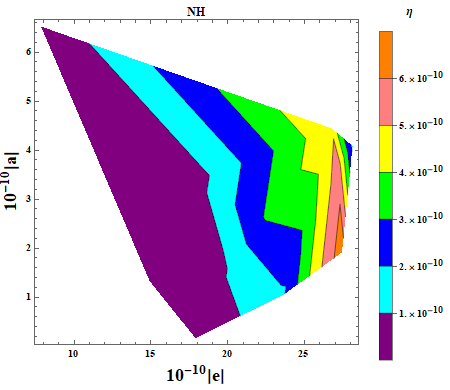}
	\end{center}
	\begin{center}
		\caption{Variation of model parameters(in eV) with the baryon asymmetry as contour for normal hierarchy.Planck limits on BAU $\eta = 6.1\times10^{-10}$.}
		\label{fig1}
	\end{center}
\end{figure}

\begin{figure}[H]
	\begin{center}
		\includegraphics[width=0.30\textwidth]{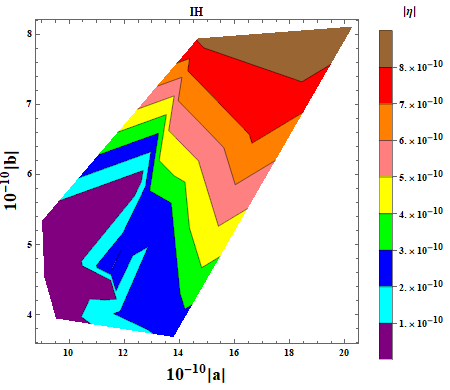}
		\includegraphics[width=0.30\textwidth]{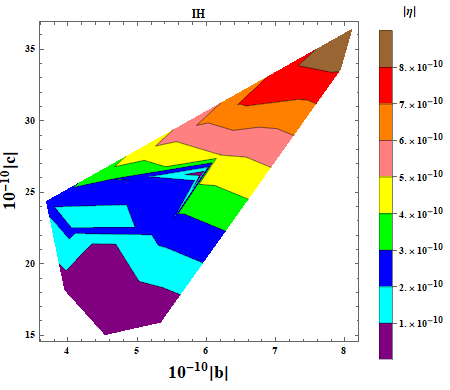}\\ 
		\includegraphics[width=0.30\textwidth]{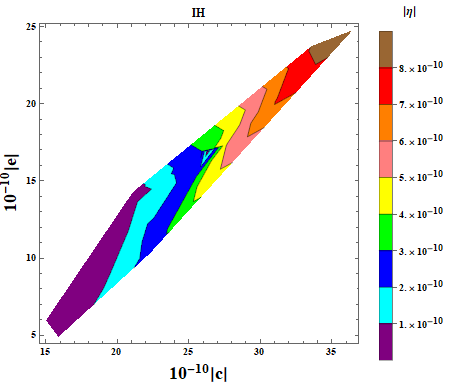}
		\includegraphics[width=0.30\textwidth]{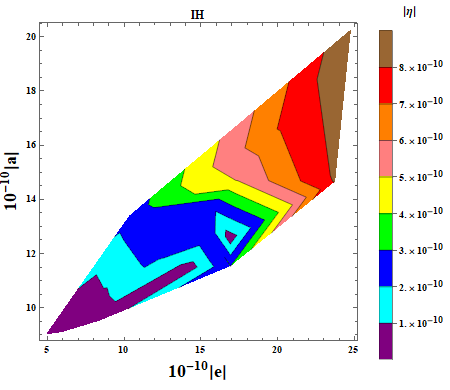}
	\end{center}
	\begin{center}
		\caption{Variation of model parameters(in eV) with the baryon asymmetry as contour for inverted hierarchy.Planck limits on BAU $\eta = 6.1\times10^{-10}$.}
		\label{fig2}
	\end{center}
\end{figure}

\begin{figure}[H]
	\begin{center}
		\includegraphics[width=0.45\textwidth]{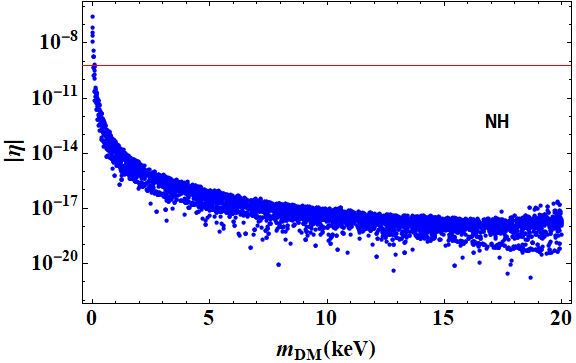}
		\includegraphics[width=0.45\textwidth]{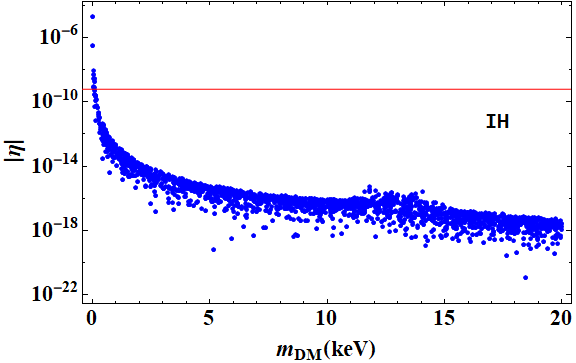}
	\end{center}
	\begin{center}
		\caption{BAU as a function of dark matter mass for NH and IH. The horizontal red line represents the Planck limits on BAU $\eta = 6.1\times10^{-10}$.}
		\label{fig3}
	\end{center}
\end{figure}

\begin{figure}[H]
	\begin{center}
		\includegraphics[width=0.35\textwidth]{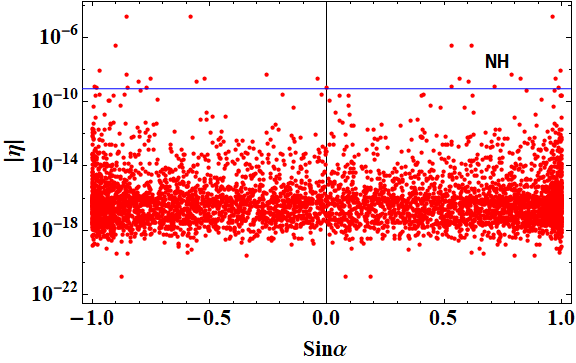}
		\includegraphics[width=0.35\textwidth]{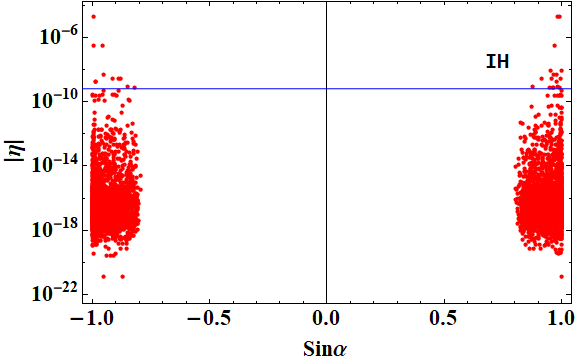}\\
		\includegraphics[width=0.35\textwidth]{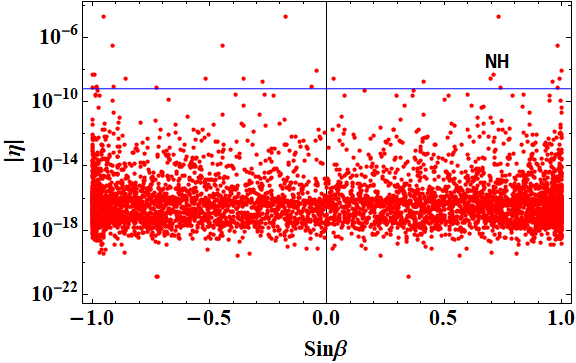} 
		\includegraphics[width=0.35\textwidth]{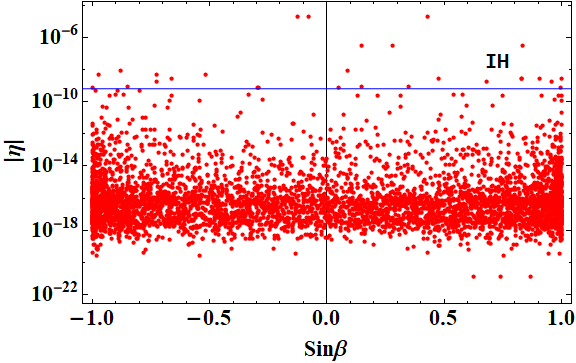}
	\end{center}
	\begin{center}
		\caption{Baryon asymmetry as a function of Majorana CP phases for normal and inverted hierarchy.The horizontal blue line represents the Planck limits on BAU $\eta = 6.1\times10^{-10}$.}
		\label{fig4}
	\end{center}
\end{figure}

\begin{figure}[H]
	\begin{center}
		\includegraphics[width=0.40\textwidth]{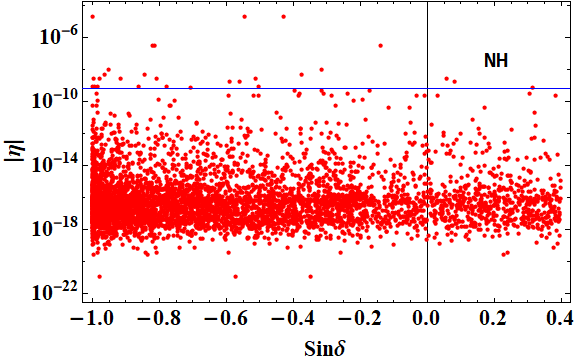}
		\includegraphics[width=0.40\textwidth]{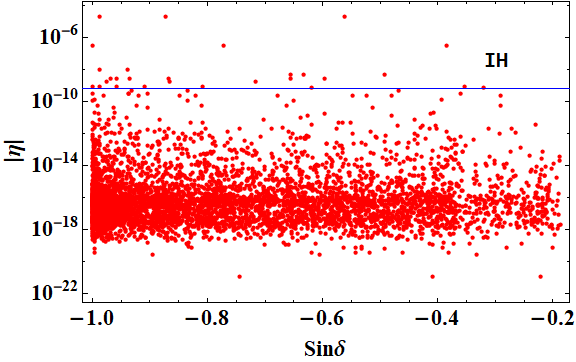}
		
	\end{center}
	\begin{center}
		\caption{Baryon asymmetry as a function of Dirac CP phase for normal and inverted hierarchy. The horizontal blue line represents the Planck limits on BAU $\eta = 6.1\times10^{-10}$}
		\label{fig5}
	\end{center}
\end{figure}

\begin{figure}[H]
	\begin{center}
		\includegraphics[width=0.45\textwidth]{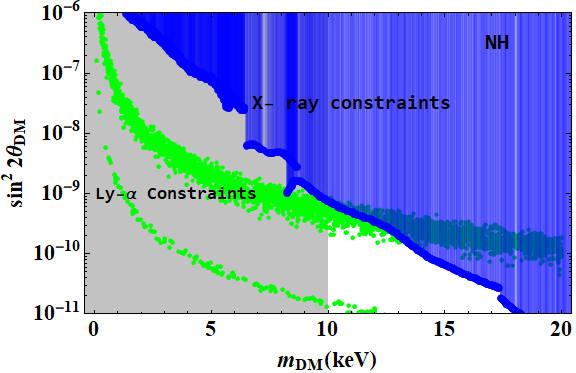}
		\includegraphics[width=0.45\textwidth]{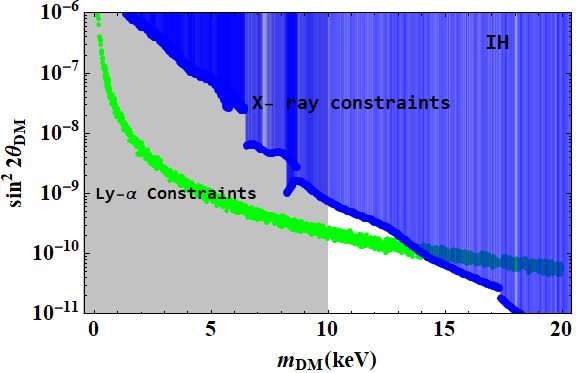}
	\end{center}
	\begin{center}
		\caption{DM-active mixing as a function of the mass of the DM for both normal hierarchy as well as inverted hierarchy. The cosmological limits are imposed in the figure.}
		\label{fig6}
	\end{center}
\end{figure}

\begin{figure}[H]
	\begin{center}
		\includegraphics[width=0.40\textwidth]{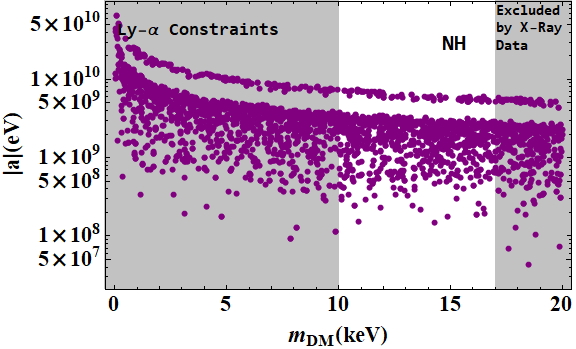}
		\includegraphics[width=0.40\textwidth]{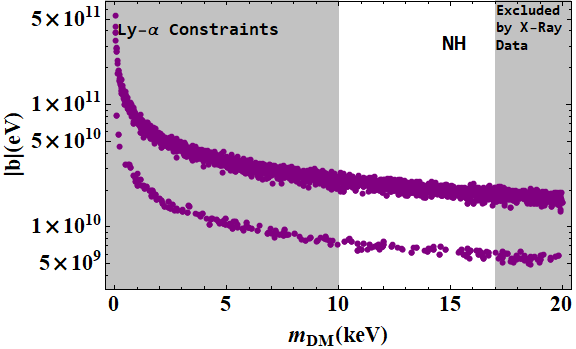}\\ 
		\includegraphics[width=0.40\textwidth]{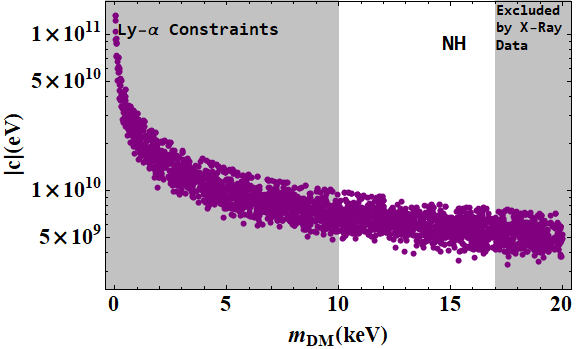}
		\includegraphics[width=0.40\textwidth]{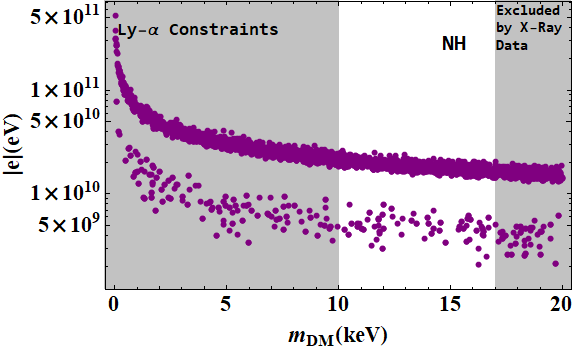}
	\end{center}
	\begin{center}
		\caption{Dark Matter mass as a function of the model parameters in normal hierarchy. The shaded regions are excluded by X-ray and $Ly-\alpha$ data.}
		\label{fig7}
	\end{center}
\end{figure}

\begin{figure}[H]
	\begin{center}
		\includegraphics[width=0.38\textwidth]{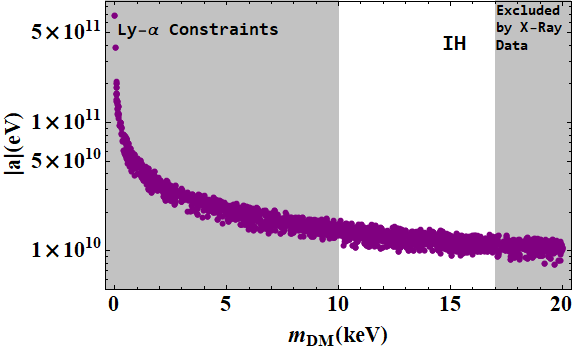}
		\includegraphics[width=0.38\textwidth]{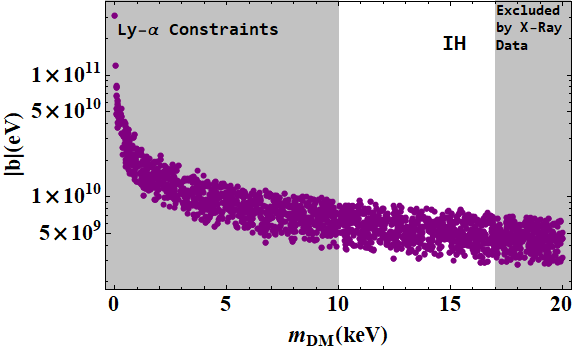}\\ 
		\includegraphics[width=0.38\textwidth]{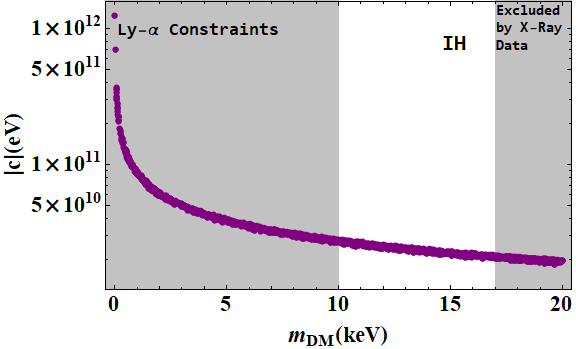}
		\includegraphics[width=0.38\textwidth]{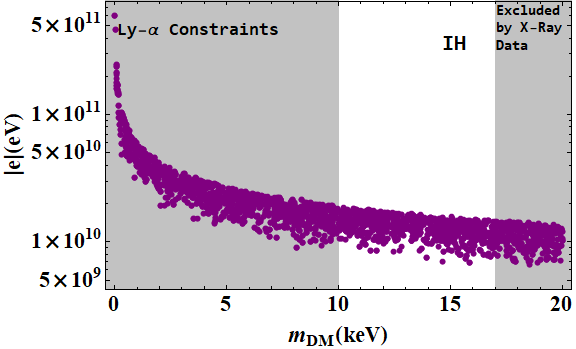}
	\end{center}
	\begin{center}
		\caption{Dark Matter mass as a function of the model parameters for inverted hierarchy. The shaded regions are excluded by X-ray and $Ly-\alpha$ data.}
		\label{fig8}
	\end{center}
\end{figure}

\begin{figure}[H]
	\begin{center}
		\includegraphics[width=0.43\textwidth]{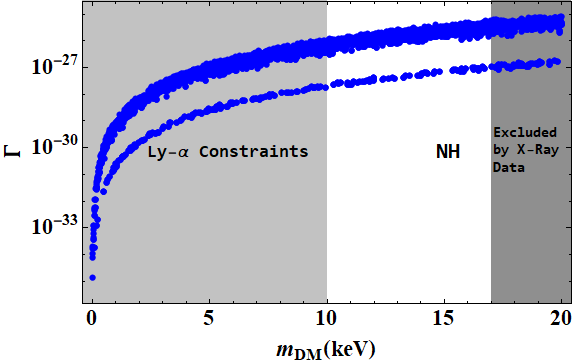}		
		\includegraphics[width=0.43\textwidth]{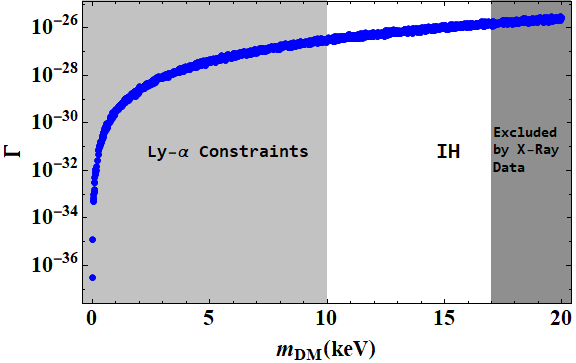}
	\end{center}
	\begin{center}
		\caption{Decay rate (in $s^{-1}$)of the lightest sterile neutrino as a function of DM mass for normal hierarchy as well as inverted hierarchy.}
		\label{fig9}
	\end{center}
\end{figure}

\begin{figure}[H]
	\begin{center}
		\includegraphics[width=0.4\textwidth]{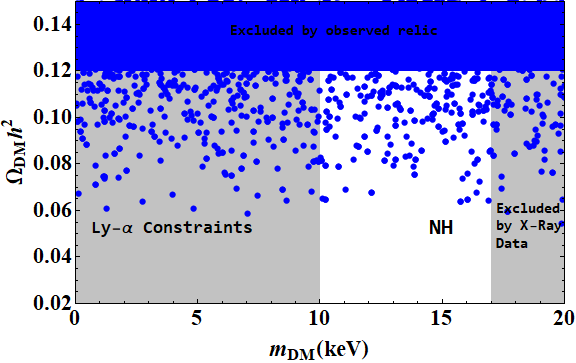}
		\includegraphics[width=0.4\textwidth]{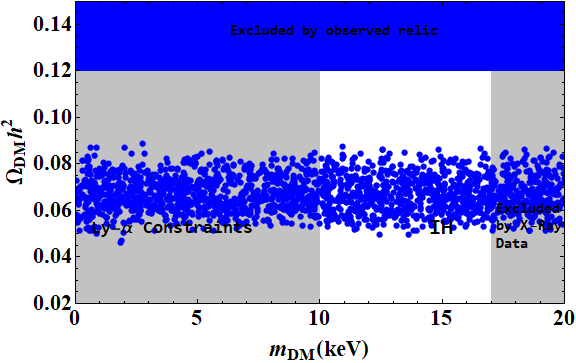}
	\end{center}
	\begin{center}
		\caption{Sterile neutrino contribution to DM abundance in NH and IH.}
		\label{fig10}
	\end{center}
\end{figure}

\begin{figure}[H]
	\begin{center}
		\includegraphics[width=0.38\textwidth]{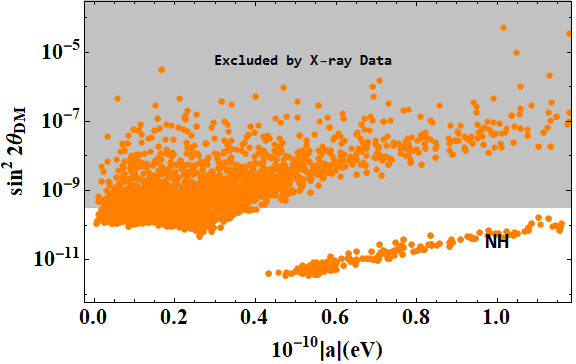}
		\includegraphics[width=0.38\textwidth]{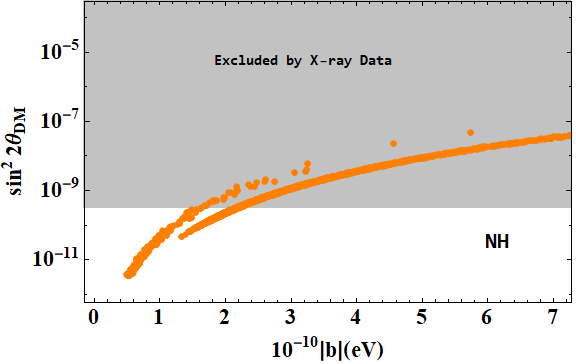}\\ 
		\includegraphics[width=0.38\textwidth]{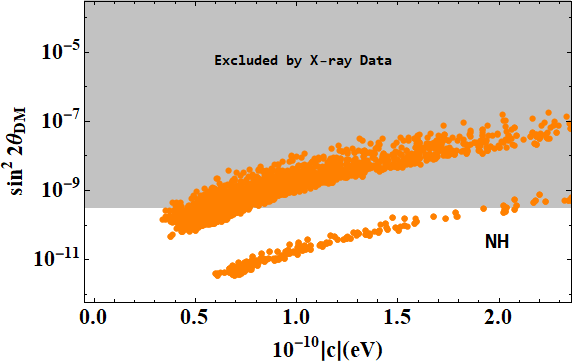}
		\includegraphics[width=0.38\textwidth]{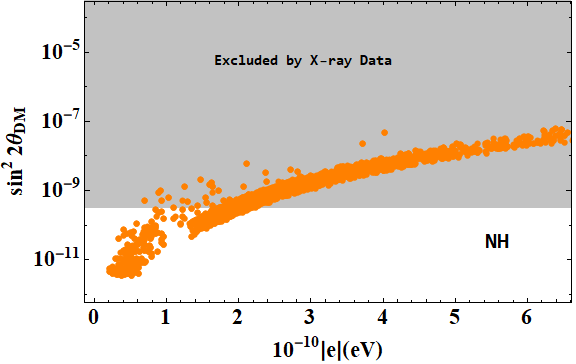}
	\end{center}
	\begin{center}
		\caption{DM-active mixing as a function of the model parameters in normal hierarchy. The shaded regions are excluded by X-ray and $Ly-\alpha$ data.}
		\label{fig11}
	\end{center}
\end{figure}

\begin{figure}[H]
	\begin{center}
		\includegraphics[width=0.40\textwidth]{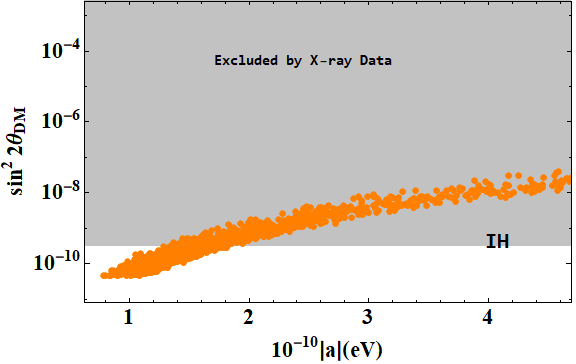}
		\includegraphics[width=0.40\textwidth]{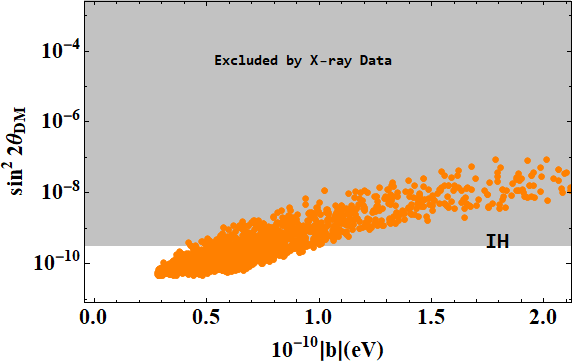}\\ 
		\includegraphics[width=0.40\textwidth]{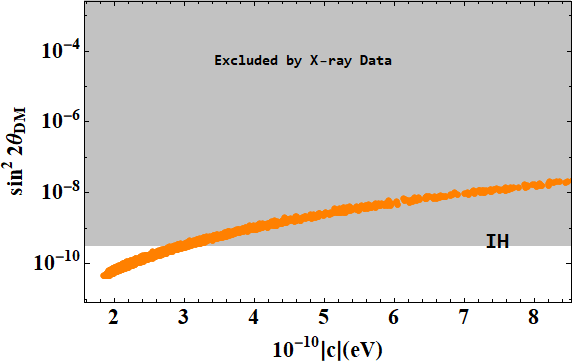}
		\includegraphics[width=0.40\textwidth]{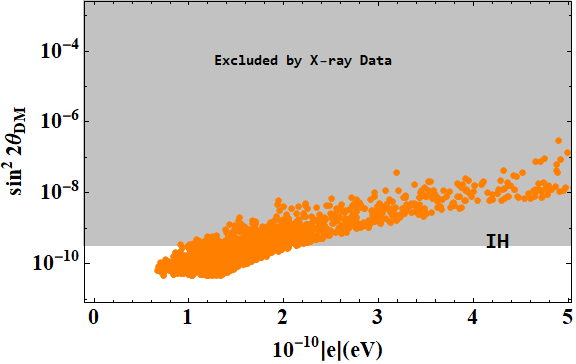}
	\end{center}
	\begin{center}
		\caption{DM-active mixing as a function of the model parameters for inverted hierarchy. The shaded regions are excluded by X-ray and $Ly-\alpha$ data.}
		\label{fig12}
	\end{center}
\end{figure}

The mass of sterile neutrino dark matter can be obtained by diagonalising $M_{H}$ and DM-active mixing can be determined using \eqref{eq:b}. Then we study the DM phenomenology which involves the calculation of the decay rates of the sterile DM using\eqref{eq:d}, DM-active mixing, and also the relic abundance of the proposed candidate for normal hierarchy(NH) as well as inverted hierarchy(IH). All of these are dependent on our model parameters as mentioned earlier. Hence, the same set of model parameters that are supposed to produce correct neutrino phenomenology can also be used to estimate the DM-active mixing, relic abundance, and the decay rate of the sterile neutrino dark matter. 
\begin{itemize}
	\item {We have shown different plots obtained from our numerical analysis carried out for normal hierarchy (NH) as well as inverted hierarchy (IH) from fig \ref{fig1} to fig \ref{fig12}.}
	\item {Fig \ref{fig1} and Fig \ref{fig2} represent two parameter contour plots with the baryon asymmetry as a contour for NH and IH respectively when the lightest sterile neutrino is in keV range. The values of the model parameters giving rise to the observed BAU are given in table \ref{tab5}.}
	\item {Fig \ref{fig3} shows the variation of baryon asymmetry as a function of DM mass for normal as well as inverted hierarchy. These two observables are cosmological and our model satisfies the existing limits on both the observables. It is evident from the plots that the observed baryon asymmetry is satisfied for the dark matter mass around 3 keV for NH as well as IH.}
	\item {Fig \ref{fig4} shows the variation of baryon asymmetry as a function of Majorana CP phases for normal as well as inverted hierarchy. It is noticeable that our model highly constrains the Majorana CP phase $\alpha$ which does not have any prior conclusive range. However, our model discards the sin $\alpha$ range from -0.2 to 0.2 in case of IH.}
	\item {The variation of baryon asymmetry as a function of Dirac CP phases for normal as well as inverted hierarchy are shown in fig \ref{fig5}. It can be seen from this plot that it cannot give preference for any particular value of the Dirac CP phase.}
	\item {Fig \ref{fig6} to fig \ref{fig10} are the plots for dark matter phenomenology when the sterile neutrino is in keV range and behaves as a viable DM candidate.} 
	\item {Fig \ref{fig6} represents the two dimensional parameter space for DM mass and DM-active for both NH and IH. The limits on the mass and active-sterile mixing from the requirement of a good DM candidate are ($0.4-50$)keV and ($10^{-12}-10^{-8}$) respectively. We have incorporated the cosmological bounds from Lyman-$\alpha$ and X-ray data in the figure. The figure indicates that the allowed dark matter mass lies within the range of ($10-14$) keV for NH and ($10-17$) keV for IH.}
	\item {Fig \ref{fig7} and fig \ref{fig8} show the variation of DM mass as a function of the model parameters for normal hierarchy and inverted hierarchy respectively. Cosmological bounds from Lyman-$\alpha$ and X-ray data are implemented in the figure. X-ray limits on fig \ref{fig6} exclude the masses above 17 keV and that limits are also included in the fig \ref{fig7} and fig \ref{fig8}.}
	\item {We have shown the prediction of the decay rate of the lightest sterile neutrino as a function of DM mass for both the mass hierarchies in fig \ref{fig9}. It is observed that the decay rate is negligible and lies within the range $10^{-36}$ to $10^{-27}$ $s^{-1}$  for both the hierarchies. The constraints from structure formation are also imposed in the figure.}
	\item {Fig \ref{fig10} shows the relic abundance of the proposed DM candidate as a function of DM mass. It is noticeable that both the hierarchical patterns of the neutrino mass lead to an overabundance of sterile neutrino dark matter which region is however excluded in our analysis. For the inverted hierarchical pattern, we have not obtained good results satisfying the relic abundance while the normal hierarchical pattern shows better predictions on relic abundance.}
	\item {Fig \ref{fig11} and fig \ref{fig12} represent DM-active mixing as a function of the model parameters for normal hierarchy and inverted hierarchy respectively. We have imposed the cosmological bounds from Lyman-$\alpha$ and X-ray data in the figure. The regions excluded by the X-ray limits on fig \ref{fig6} are also implemented here. A wide range of all the model parameters agree with the limits of DM-active mixing in our model.}
\end{itemize}

\begin{table}[H]
	\centering
	\begin{tabular}{|c|c|c|}
		\hline 
		Model Parameter  & NH(eV) & IH(eV) \\ 
		\hline 
		a & $2\times 10^{10}-3.5\times 10^{10}$ &  $ 1.6\times 10^{11}-1.9\times 10^{11}$\\
		\hline
		b & $2.8\times 10^{11}-3.1\times 10^{11}$ &  $7.0\times 10^{10}-7.8\times 10^{10}$\\
		\hline 
		c & $6\times 10^{10}-7\times10^{10}$ &  $2.9\times 10^{11}-3.4\times 10^{11}$ \\
		\hline 
		e & $2.5\times 10^{11}-2.9\times 10^{11}$ &  $2\times 10^{11}-2.6\times 10^{11}$\\
		\hline
	\end{tabular} 
	\caption{Allowed range of the model parameters satisfying observed limit of BAU.} \label{tab5}
\end{table} 

\section{\label{sec:level7}Discussion and Conclusion}
In this work, we have addressed neutrino mass and baryon asymmetry of the universe along with the proposal of a viable dark matter candidate in an inverse seesaw framework. The standard model is extended by discrete flavor symmetry $S_{4}\times Z_{4}\times Z_{3}$. The field contents and their transformations under  $S_{4}$ symmetry are so chosen that these fields lead to a light neutrino mass matrix $ m_{\nu}$ which is consistent with the broken $ \mu-\tau $ symmetry. In this model, the lepton asymmetry is generated by the CP violating decay of the lightest quasi-Dirac pair which then converts to baryon asymmetry through the sphaleron process. This model has the special feature that there exists a sterile state in the keV scale, which can contribute to significant dark matter phenomenology.

After constructing the model, the model parameters have been evaluated by comparing the light neutrino mass matrix arising from the model with the one constructed from light neutrino parameters for both normal as well as inverted hierarchy. We then feed the model parameters in the calculation of DM-active mixing and the relic abundance considering the lightest sterile neutrino as a dark matter candidate. The stability of the proposed dark matter has been verified by calculating the decay rate of the lightest sterile neutrino in the process $N\longrightarrow\nu+\gamma$ (N being the sterile neutrino) with the same range of allowed parameters. This results in a negligible decay rate ensuring the stability of the dark matter candidate at least in the cosmological scales. The parameter space is confined due to the cosmological bounds from X-ray and Ly-$\alpha$ data which have been implemented in our analysis. The parameter space is compatible with XQ-100 Ly-$\alpha$ data along with the X-ray bounds. However, the current Ly-$\alpha$ constraint using XQ-100 + BOSS data on non resonantly produced sterile neutrino dark matter is in tension with the upper limit of mass provided by the X-ray data. This conflict may arise due to the fact that the dark matter candidate may feature both resonant and non resonant component. The study on such sterile neutrino dark matter within the inverse seesaw framework can be done in future.

Apart from the calculation involved in the DM phenomenology, the same allowed range of model parameters are used to calculate the baryon asymmetry of the universe. We have considered the possibility that the lightest quasi-Dirac pair generates the lepton asymmetry, whereas the asymmetry created by the heavier quasi-Dirac pair has been washed out. It is quite interesting that a wide range of model parameters agree with the observed baryon asymmetry of the universe for both the hierarchies. We have also found the correlation between the two cosmological observables which restricts our parameter space.

The inverse seesaw model is known to be able to simultaneously account for neutrino masses,mixing, and dark matter (DM). However, the implementation of an additional discrete symmetry further constrains the model, enhancing its predictability and testability perspectives, especially concerning its flavor structure and CP properties. We have shown that the ISS augmented with an additional $S_{4}$ flavor symmetry can still account for neutrino as well as DM phenomenology with proper explanation of the observed baryon asymmetry of the universe which is a significant feature of this work. Thus besides the prediction for experimentally observed neutrino parameters, this model can explain two of the important puzzles of particle physics and cosmology, the baryon asymmetry of the universe as well as the dark matter. The proposed model of explaining the two important cosmological issues can shed light on different future and present dark matter detection experiments. This model may have some other interesting implications in collider as well as rare decay experiments like lepton flavor violation which we leave for our future study. The impacts of the model on lepton number violating processes can also be studied further.

\appendix
\section{Analytical Expressions for the CP asymmetry in terms of model parameters} 
\label{appen1}
The full Majorana mass matrix in the flavor basis $(N_{i}, s_{i})$ is given by
\[
M = \begin{pmatrix}
0 & M_{N} \\
M_{N}^{T} &\mu\\
\end{pmatrix}
\]
The Yukawa Lagrangian in this basis is given by	
\begin{equation*}
L_{y} = y_{i\alpha}\bar{N_{i}}\phi ^{\dagger} l_{\alpha} +{M_{N_{ij}}}N_{i}^{T}C^{-1}s_{j} + \dfrac{1}{2} \mu_{ij}{s_{i}^{T}C^{-1}s_{j}} + h.c
\end{equation*}

In order to calculate the CP asymmetry in this framework, it is more 
convenient to work in the basis in which the RH Majorana neutrino mass matrix is diagonal with real and positive eigenvalues. The Lagrangian in that case has the Lagrangian
\begin{equation}
L_{h} = h_{i\alpha}\bar{N_{i}}\phi l_{\alpha} + \dfrac{1}{2} M_{i}N_{i}^{T}C^{-1}N_{i} + h.c
\end{equation}
In our model, there are two quasi-Dirac pairs and the lighteset sterile neutrino decouples from them. The analytic expression for CP asymmetry can be derived in these two limits and for two sets of RH neutrinos i.e. $(N_{i}, S_{i})$ with $i = 1,2 $. Now, the $4\times4$ version of Majorana matrix becomes
\begin{equation}\label{eq:4}
M =\left(\begin{array}{cc}
0 & M_{N_{2\times2}} \\
M_{N_{2\times2}}^{T} &\mu_{2\times2}\\
\end{array}\right)
\end{equation}

The Majorana mass matrix can be reduced to a simple block diagonal form which decouples the $(N_{1},S_{1})$ and $(N_{2}, s_{2})$ sector as follows
\begin{equation}\label{eq:4a}
M = \left(\begin{array}{cccc}
0 & 0 & M_{N_1} & 0\\
0 & 0 & 0 & M_{N_2} \\
M_{N_1} & 0 & \mu_{11} & 0 \\
0 & M_{N_2} & 0 & \mu_{22} \\
\end{array}\right)\xrightarrow{r_{2}\leftrightarrow r_{3},c_{2}\leftrightarrow c_{3}}
\left(\begin{array}{cccc}
0 & M_{N_1} & 0 & 0\\
M_{N_1} & \mu_{11} & 0 & 0\\
0 & 0 & 0 & M_{N_2} \\
0 & 0 &  M_{N_2} & \mu_{22} \\
\end{array}\right)
\end{equation}
Then in the $(N_{i}, s_{i})$ flavor basis, we have the $2\times2$ matrices
\begin{equation}\label{eq:5}
M_{i} = \left(\begin{array}{cc}
0 & M_{N_i} \\
M_{N_i}&\mu_{ii} \\
\end{array}\right) =
\left(\begin{array}{cc}
0 & M_{N_i} \\
M_{N_i}& \epsilon_{i}M_{N_i}e^{i\theta_{i}} \\
\end{array}\right)
\end{equation}
where $\varepsilon_{i} = \frac{\mu_{ii}}{M_{N_i}}\ll 1$ and The $M_{i}$ is diagonalized with real and positive eigenvalues by a unitary transformation $ U_{i}^{T}M_{i}U_{i} $ where
\begin{equation}\label{eq:57}
U_{i} = \left(\begin{array}{cc}
-icos\alpha_{i}e^{i\theta_{i}/2} & sin\alpha_{i}e^{i\theta_{i}/2} \\
isin\alpha_{i}e^{-i\theta_{i}/2}& cos\alpha_{i}e^{-i\theta_{i}/2} \\
\end{array}\right)
\end{equation}
and the mixing angles are given by
\begin{equation}
cos\alpha_{i} \simeq \frac{1}{\sqrt{2}}(1+ \dfrac{\varepsilon_{i}}{4}),
sin\alpha_{i} \simeq \frac{1}{\sqrt{2}}(1- \dfrac{\varepsilon_{i}}{4})
\end{equation}
The corresponding mass eigenvalues are given by
\begin{equation*}
M_{i} \simeq M_{N_{i}}(1\pm \dfrac{\varepsilon_{i}}{2}), (i=1,2 ; j=1,2,3,4)
\end{equation*}
From the expression,it is evident that the mass splitting within a quasi -Dirac pair is given by $\mu_{ii} $

The Yukawa couplings in this diagonal mass basis are related to the couplings in the flavor basis as follows:

\begin{equation*}
h_{1\alpha} \simeq \frac{ie^{-i\theta_{1}}}{\sqrt{2}}(1+ \dfrac{\epsilon_{1}}{4})y_{1\alpha}
\end{equation*}
\begin{equation*}
h_{2\alpha} \simeq \frac{e^{-i\theta_{1}}}{\sqrt{2}}(1- \dfrac{\epsilon_{1}}{4})y_{1\alpha}
\end{equation*}
\begin{equation*}
h_{3\alpha} \simeq \frac{ie^{-i\theta_{2}}}{\sqrt{2}}(1+ \dfrac{\epsilon_{2}}{4})y_{2\alpha}
\end{equation*}
\begin{equation*}
h_{4\alpha} \simeq \frac{e^{-i\theta_{2}}}{\sqrt{2}}(1- \dfrac{\epsilon_{2}}{4})y_{2\alpha}
\end{equation*}
The CP-asymmetry for the decay of one of the quasi-Dirac
particles can be calculated using the expression
\begin{equation*}
\epsilon_{i} = \frac{1}{8\pi}\sum \dfrac{Im[(hh\dagger)^{2}_{ij}]}{\sum|h_{i\beta}|^{2}}f_{ij^{v}}
\end{equation*}
We have considered the CP asymmetry generated by the first pair . Now say for i= 1 and i=2 The CP-asymmetry parameter will be
\begin{equation*}
\epsilon_{1} = \frac{1}{8\pi}\sum_{j\neq1} \dfrac{Im[(hh\dagger)^{2}_{1j}]}{\sum|h_{1\beta}|^{2}}f_{1j^{v}} \simeq \frac{\varepsilon_{2}}{16\pi \sum|y_{1\beta}|^{2}}Im [e^{i(\theta_{1}-\theta_{2})}(\sum_{\beta} y_{1\alpha}^{\ast}y_{2\alpha})^{2}]f_{13}^{v}
\end{equation*}
\begin{equation*}
\epsilon_{2} = \frac{1}{8\pi}\sum_{j\neq2} \dfrac{Im[(hh\dagger)^{2}_{2j}]}{\sum_{\beta}|h_{1\beta}|^{2}}f_{2j^{v}} \simeq \frac{\varepsilon_{2}}{16\pi \sum|y_{1\beta}|^{2}}Im [ie^{i(\theta_{1}-\theta_{2})}(\sum y_{1\alpha}^{\ast}y_{2\alpha})^{2}]f_{23}^{v}
\end{equation*}
Here, $\varepsilon_{1}= \frac{p}{f}$ and  $\varepsilon_{2}= \frac{p}{g}$
and it is assumed that $f_{13} = f_{14}$. Here, $j = 2$ term vanishes as there is no imaginary part in that case. 
In our model  $M_{N_{1}} = f$ , $M_{N_{2}} = g$ ,$\mu_{11} = \mu_{22}= p$ and thus it leads to the diagonalising matrix of M in our model as
\begin{equation*}
U_{1}=\left(\begin{array}{ccc}
-\frac{1}{2f}(p+\sqrt{4f^{2}+ p^{2}}) & 1 \\
-\frac{1}{2f}(p-\sqrt{4f^{2}+ p^{2}})& 1
\end{array}\right)\
\end{equation*}
\begin{equation}\label{eq:60}
U_{2}=\left(\begin{array}{ccc}
-\frac{1}{2g}(p+\sqrt{4g^{2}+ p^{2}}) & 1 \\
-\frac{1}{2g}(p-\sqrt{4g^{2}+ p^{2}})& 1
\end{array}\right)\
\end{equation}
with
\begin{equation*}
cos\alpha_{1} \simeq \frac{1}{\sqrt{2}}(1+ \frac{p}{4f}),
sin\alpha_{1} \simeq \frac{1}{\sqrt{2}}(1- \frac{p}{4f})
\end{equation*}
\begin{equation}
cos\alpha_{2} \simeq \frac{1}{\sqrt{2}}(1+ \frac{p}{4g}),
sin\alpha_{2} \simeq \frac{1}{\sqrt{2}}(1- \frac{p}{4g})
\end{equation}
Comparing these matrices arising from our model with the one mentioned in equation \ref{eq:57} and with further simplifications we obtain the expressions for $e^{i(\theta_{1}-\theta_{2})}$ as,
\begin{equation}
e^{i(\theta_{1}-\theta_{2})} = \frac{32fg}{(4g+p)(4f-p)}
\end{equation}
Again, the decay width of one of the quasi-Dirac pairs (say i=1) can be written as,
\begin{equation}\label{54}
\Gamma_{1} = \frac{M_{1}}{8\pi}(hh^{\dagger})_{11} = \frac{1}{2}(1+\frac{\varepsilon_{1}}{2})\sum_{\alpha}y_{1\alpha}^{*}y_{1\alpha}
\end{equation} 
\section*{Acknowledgements}
NG acknowledges Department of Science and Technology (DST),India(grant DST/INSPIRE Fellowship/2016/IF160994) for the financial assistantship. The work of MKD is supported by the Department of Science and Technology, Government of India under the project no. $EMR/2017/001436$.
\bibliographystyle{paper}
\bibliography{Lepto}

\begin{thebibliography}{10}

\bibitem{mass-mixingneutrino}
King, S.~F.
\newblock Neutrino mass models.
\newblock \emph{Reports on Progress in Physics}, 67(2):107, 2003

\bibitem{king2014neutrino}
King, S.~F., Merle, A., Morisi, S., Shimizu, Y., and Tanimoto, M.
\newblock Neutrino mass and mixing: from theory to experiment.
\newblock \emph{New Journal of Physics}, 16(4):045018, 2014

\bibitem{rubin1970rotation}
Rubin, V.~C. and Ford~Jr, W.~K.
\newblock Rotation of the Andromeda nebula from a spectroscopic survey of
  emission regions.
\newblock \emph{The Astrophysical Journal}, 159:379, 1970

\bibitem{clowe2006direct}
Clowe, D., Brada{\v{c}}, M., Gonzalez, A.~H., Markevitch, M., Randall, S.~W.
  et~al.
\newblock A direct empirical proof of the existence of dark matter.
\newblock \emph{The Astrophysical Journal Letters}, 648(2):L109, 2006

\bibitem{ade2016ade}
Ade, P.
\newblock PAR Ade et al.(Planck Collaboration), Astron. Astrophys. 594, A13
  (2016).
\newblock \emph{Astron. Astrophys.}, 594:A13, 2016

\bibitem{taoso2008dark}
Taoso, M., Bertone, G., and Masiero, A.
\newblock Dark matter candidates: a ten-point test.
\newblock \emph{Journal of Cosmology and Astroparticle Physics}, 2008(03):022,
  2008

\bibitem{mukherjee2017common}
Mukherjee, A., Borah, D., and Das, M.~K.
\newblock Common origin of nonzero $\theta$ 13 and dark matter in an S 4 flavor
  symmetric model with inverse seesaw mechanism.
\newblock \emph{Physical Review D}, 96(1):015014, 2017

\bibitem{adhikari2017white}
Adhikari, R., Agostini, M., Ky, N.~A., Araki, T., Archidiacono, M. et~al.
\newblock A white paper on keV sterile neutrino dark matter.
\newblock \emph{Journal of cosmology and astroparticle physics}, 2017(01):025,
  2017

\bibitem{kusenko2009sterile}
Kusenko, A.
\newblock Sterile neutrinos: the dark side of the light fermions.
\newblock \emph{Physics Reports}, 481(1-2):1--28, 2009

\bibitem{eVsterile}
Hamann, J., Hannestad, S., Raffelt, G.~G., and Wong, Y.~Y.
\newblock Sterile neutrinos with eV masses in cosmology—how disfavoured
  exactly?
\newblock \emph{Journal of Cosmology and Astroparticle Physics}, 2011(09):034,
  2011

\bibitem{admixturesterile}
Dodelson, S. and Widrow, L.~M.
\newblock Sterile neutrinos as dark matter.
\newblock \emph{Physical Review Letters}, 72(1):17, 1994

\bibitem{barry2011light}
Barry, J., Rodejohann, W., and Zhang, H.
\newblock Light sterile neutrinos: models and phenomenology.
\newblock \emph{Journal of High Energy Physics}, 2011(7):91, 2011

\bibitem{PhysRevD.98.030001}
Tanabashi, M., Hagiwara, K., Hikasa, K., Nakamura, K., Sumino, Y. et~al.
\newblock Review of Particle Physics.
\newblock \emph{Phys. Rev. D}, 98:030001, 2018

\bibitem{Sakharov:1967dj}
Sakharov, A.~D.
\newblock {Violation of CP Invariance, C asymmetry, and baryon asymmetry of the
  universe}.
\newblock \emph{Pisma Zh. Eksp. Teor. Fiz.}, 5:32--35, 1967.
\newblock [Usp. Fiz. Nauk161,no.5,61(1991)]

\bibitem{Fukugita:1986hr}
Fukugita, M. and Yanagida, T.
\newblock {Baryogenesis Without Grand Unification}.
\newblock \emph{Phys. Lett.}, B174:45--47, 1986

\bibitem{PILAFTSIS_1999}
PILAFTSIS, A.
\newblock HEAVY MAJORANA NEUTRINOS AND BARYOGENESIS.
\newblock \emph{International Journal of Modern Physics A}, 14(12):1811–1857,
  1999.
\newblock ISSN 1793-656X

\bibitem{kolb2018early}
Kolb, E.
\newblock \emph{The early universe}.
\newblock CRC Press, 2018

\bibitem{an2012observation}
An, F., Bai, J., Balantekin, A., Band, H., Beavis, D. et~al.
\newblock Observation of electron-antineutrino disappearance at Daya Bay.
\newblock \emph{Physical Review Letters}, 108(17):171803, 2012

\bibitem{arhrib2010collider}
Arhrib, A., Bajc, B., Ghosh, D.~K., Han, T., Huang, G.-Y. et~al.
\newblock Collider signatures for the heavy lepton triplet in the type I+ III
  seesaw mechanism.
\newblock \emph{Physical Review D}, 82(5):053004, 2010

\bibitem{foot1989see}
Foot, R., Lew, H., He, X.-G., and Joshi, G.~C.
\newblock See-saw neutrino masses induced by a triplet of leptons.
\newblock \emph{Zeitschrift f{\"u}r Physik C Particles and Fields},
  44(3):441--444, 1989

\bibitem{mohapatra1981neutrino}
Mohapatra, R.~N. and Senjanovi{\'c}, G.
\newblock Neutrino masses and mixings in gauge models with spontaneous parity
  violation.
\newblock \emph{Physical Review D}, 23(1):165, 1981

\bibitem{Abada2014}
Abada, A. and Lucente, M.
\newblock Looking for the minimal inverse seesaw realisation.
\newblock \emph{Nuclear Physics B}, 885:651--678, 2014

\bibitem{abada2017neutrino}
Abada, A., Arcadi, G., Domcke, V., and Lucente, M.
\newblock Neutrino masses, leptogenesis and dark matter from small lepton
  number violation?
\newblock \emph{Journal of Cosmology and Astroparticle Physics}, 2017(12):024,
  2017

\bibitem{gautam2019phenomenology}
Gautam, N. and Das, M.~K.
\newblock Phenomenology of keV scale sterile neutrino dark matter with $S_{4}$
  flavor symmetry, 2019

\bibitem{Krishnan:2012me}
Krishnan, R., Harrison, P., and Scott, W.
\newblock {Simplest Neutrino Mixing from S4 Symmetry}.
\newblock \emph{JHEP}, 04:087, 2013

\bibitem{awasthi2013neutrinoless}
Awasthi, R.~L., Parida, M., and Patra, S.
\newblock Neutrinoless double beta decay and pseudo-Dirac neutrino mass
  predictions through inverse seesaw mechanism.
\newblock \emph{arXiv preprint arXiv:1301.4784}, 2013

\bibitem{lindner2014neutrino}
Lindner, M., Schmidt, S., and Smirnov, J.
\newblock Neutrino masses and conformal electro-weak symmetry breaking.
\newblock \emph{Journal of High Energy Physics}, 2014(10):177, 2014

\bibitem{abada2014dark}
Abada, A., Arcadi, G., and Lucente, M.
\newblock Dark Matter in the minimal Inverse Seesaw mechanism.
\newblock \emph{Journal of Cosmology and Astroparticle Physics}, 2014(10):001,
  2014

\bibitem{Lucente:2018uaj}
Lucente, M., Abada, A., Arcadi, G., Domcke, V., Drewes, M. et~al.
\newblock {Freeze-in leptogenesis with 3 right-handed neutrinos}.
\newblock \emph{PoS}, ICHEP2018:306, 2019

\bibitem{Hambye_2012}
Hambye, T.
\newblock Leptogenesis: beyond the minimal type I seesaw scenario.
\newblock \emph{New Journal of Physics}, 14(12):125014, 2012.
\newblock ISSN 1367-2630

\bibitem{Choubey_2010}
Choubey, S., King, S.~F., and Mitra, M.
\newblock Vanishing of theCPasymmetry in leptogenesis due to form dominance.
\newblock \emph{Physical Review D}, 82(3), 2010.
\newblock ISSN 1550-2368

\bibitem{Agashe:2018cuf}
Agashe, K., Du, P., Ekhterachian, M., Fong, C.~S., Hong, S. et~al.
\newblock {Natural Seesaw and Leptogenesis from Hybrid of High-Scale Type I and
  TeV-Scale Inverse}.
\newblock \emph{JHEP}, 04:029, 2019

\bibitem{dev2010tev}
Dev, P.~B. and Mohapatra, R.
\newblock TeV scale inverse seesaw model in S O (10) and leptonic nonunitarity
  effects.
\newblock \emph{Physical Review D}, 81(1):013001, 2010

\bibitem{borah2018common}
Borah, D., Das, M.~K., and Mukherjee, A.
\newblock Common origin of nonzero $\theta$ 13 and baryon asymmetry of the
  Universe in a TeV scale seesaw model with A 4 flavor symmetry.
\newblock \emph{Physical Review D}, 97(11):115009, 2018

\bibitem{Covi_1996}
Covi, L., Roulet, E., and Vissani, F.
\newblock CP violating decays in leptogenesis scenarios.
\newblock \emph{Physics Letters B}, 384(1-4):169–174, 1996.
\newblock ISSN 0370-2693

\bibitem{Dev_2015}
Dev, P. S.~B., Millington, P., Pilaftsis, A., and Teresi, D.
\newblock Flavour effects in Resonant Leptogenesis from semi-classical and
  Kadanoff-Baym approaches.
\newblock \emph{Journal of Physics: Conference Series}, 631:012087, 2015.
\newblock ISSN 1742-6596

\bibitem{Blanchet_2010}
Blanchet, S., Dev, P. S.~B., and Mohapatra, R.~N.
\newblock Leptogenesis with TeV-scale inverse seesaw model inSO(10).
\newblock \emph{Physical Review D}, 82(11), 2010.
\newblock ISSN 1550-2368

\bibitem{mukherjee2018normal}
Mukherjee, A., Das, M.~K., and Sarma, J.~K.
\newblock Normal hierarchy neutrino mass model revisited with leptogenesis,
  2018

\bibitem{Pilaftsis_2005}
Pilaftsis, A. and Underwood, T. E.~J.
\newblock Electroweak-scale resonant leptogenesis.
\newblock \emph{Physical Review D}, 72(11), 2005.
\newblock ISSN 1550-2368

\bibitem{Pilaftsis_1997}
Pilaftsis, A.
\newblock CPviolation and baryogenesis due to heavy Majorana neutrinos.
\newblock \emph{Physical Review D}, 56(9):5431–5451, 1997.
\newblock ISSN 1089-4918

\bibitem{Giudice_2004}
Giudice, G., Notari, A., Raidal, M., Riotto, A., and Strumia, A.
\newblock Towards a complete theory of thermal leptogenesis in the SM and MSSM.
\newblock \emph{Nuclear Physics B}, 685(1-3):89–149, 2004.
\newblock ISSN 0550-3213

\bibitem{Blanchet:2009kk}
Blanchet, S., Hambye, T., and Josse-Michaux, F.-X.
\newblock {Reconciling leptogenesis with observable mu ---> e gamma rates}.
\newblock \emph{JHEP}, 04:023, 2010

\bibitem{lucente2016implication}
Lucente, M.
\newblock Implication of Sterile Fermions in Particle Physics and Cosmology.
\newblock \emph{arXiv preprint arXiv:1609.07081}, 2016

\bibitem{Merle2017a}
Merle, A.
\newblock keV sterile neutrino Dark Matter.
\newblock \emph{arXiv preprint arXiv:1702.08430}, 2017

\bibitem{abazajian2017sterile}
Abazajian, K.~N.
\newblock Sterile neutrinos in cosmology.
\newblock \emph{Physics Reports}, 711:1--28, 2017

\bibitem{asaka2007lightest}
Asaka, T., Shaposhnikov, M., and Laine, M.
\newblock Lightest sterile neutrino abundance within the $\nu$MSM.
\newblock \emph{Journal of High Energy Physics}, 2007(01):091, 2007

\bibitem{perez2017almost}
Perez, K., Ng, K.~C., Beacom, J.~F., Hersh, C., Horiuchi, S. et~al.
\newblock Almost closing the $\nu$ MSM sterile neutrino dark matter window with
  NuSTAR.
\newblock \emph{Physical Review D}, 95(12):123002, 2017

\bibitem{ng2019new}
Ng, K.~C., Roach, B.~M., Perez, K., Beacom, J.~F., Horiuchi, S. et~al.
\newblock New Constraints on Sterile Neutrino Dark Matter from $ NuSTAR $ M31
  Observations.
\newblock \emph{arXiv preprint arXiv:1901.01262}, 2019

\bibitem{abada2016impact}
Abada, A., De~Romeri, V., and Teixeira, A.
\newblock Impact of sterile neutrinos on nuclear-assisted cLFV processes.
\newblock \emph{Journal of High Energy Physics}, 2016(2):83, 2016

\bibitem{de2018status}
de~Salas, P., Forero, D., Ternes, C., Tortola, M., and Valle, J.
\newblock Status of neutrino oscillations 2018: 3$\sigma$ hint for normal mass
  ordering and improved CP sensitivity.
\newblock \emph{Physics Letters B}, 782:633--640, 2018

\bibitem{Aghanim:2018eyx}
Aghanim, N. et~al.
\newblock {Planck 2018 results. VI. Cosmological parameters}.
\newblock 2018

\bibitem{campos2016testing}
Campos, M.~D. and Rodejohann, W.
\newblock Testing keV sterile neutrino dark matter in future direct detection
  experiments.
\newblock \emph{Physical Review D}, 94(9):095010, 2016

\bibitem{horiuchi2014sterile}
Horiuchi, S., Humphrey, P.~J., Onorbe, J., Abazajian, K.~N., Kaplinghat, M.
  et~al.
\newblock Sterile neutrino dark matter bounds from galaxies of the Local Group.
\newblock \emph{Physical Review D}, 89(2):025017, 2014

\bibitem{boyarsky2008constraints}
Boyarsky, A., Iakubovskyi, D., Ruchayskiy, O., and Savchenko, V.
\newblock Constraints on decaying dark matter from XMM--Newton observations of
  M31.
\newblock \emph{Monthly Notices of the Royal Astronomical Society},
  387(4):1361--1373, 2008

\bibitem{abazajian2012light}
Abazajian, K.~N., Acero, M., Agarwalla, S., Aguilar-Arevalo, A., Albright, C.
  et~al.
\newblock Light sterile neutrinos: a white paper.
\newblock \emph{arXiv preprint arXiv:1204.5379}, 2012

\bibitem{meiksin2009publisher}
Meiksin, A.~A.
\newblock Publisher's Note: The physics of the intergalactic medium [Rev. Mod.
  Phys. 81, 1405 (2009)].
\newblock \emph{Reviews of Modern Physics}, 81(4):1943, 2009

\bibitem{boyarsky2009realistic}
Boyarsky, A., Lesgourgues, J., Ruchayskiy, O., and Viel, M.
\newblock Realistic sterile neutrino dark matter with keV mass does not
  contradict cosmological bounds.
\newblock \emph{Physical review letters}, 102(20):201304, 2009

\bibitem{baur2017constraints}
Baur, J., Palanque-Delabrouille, N., Yeche, C., Boyarsky, A., Ruchayskiy, O.
  et~al.
\newblock Constraints from Ly-$\alpha$ forests on non-thermal dark matter
  including resonantly-produced sterile neutrinos.
\newblock \emph{Journal of Cosmology and Astroparticle Physics}, 2017(12):013,
  2017

\bibitem{boyarsky2009lyman}
Boyarsky, A., Lesgourgues, J., Ruchayskiy, O., and Viel, M.
\newblock Lyman-$\alpha$ constraints on warm and on warm-plus-cold dark matter
  models.
\newblock \emph{Journal of Cosmology and Astroparticle Physics}, 2009(05):012,
  2009

\bibitem{Baur_2016}
Baur, J., Palanque-Delabrouille, N., Yèche, C., Magneville, C., and Viel, M.
\newblock Lyman-alpha forests cool warm dark matter.
\newblock \emph{Journal of Cosmology and Astroparticle Physics},
  2016(08):012–012, 2016.
\newblock ISSN 1475-7516

\bibitem{Y_che_2017}
Yèche, C., Palanque-Delabrouille, N., Baur, J., and Bourboux, H. d. M.~d.
\newblock Constraints on neutrino masses from Lyman-alpha forest power spectrum
  with BOSS and XQ-100.
\newblock \emph{Journal of Cosmology and Astroparticle Physics},
  2017(06):047–047, 2017.
\newblock ISSN 1475-7516

\bibitem{giganti2017neutrino}
Giganti, C., Lavignac, S., and Zito, M.
\newblock Neutrino oscillations: the rise of the PMNS paradigm.
\newblock \emph{Progress in Particle and Nuclear Physics}, 2017

\bibitem{Nath_2017}
Nath, N., Ghosh, M., Goswami, S., and Gupta, S.
\newblock Phenomenological study of extended seesaw model for light sterile
  neutrino.
\newblock \emph{Journal of High Energy Physics}, 2017(3), 2017.
\newblock ISSN 1029-8479

\end{thebibliography}
\end{document}